\documentclass[sigconf]{acmart}

\usepackage{graphicx}
\usepackage{subcaption}
\usepackage{multirow}
\usepackage{booktabs}
\usepackage[table]{xcolor}
\usepackage{placeins}
\usepackage[normalem]{ulem}
\usepackage{enumitem}
\usepackage{wrapfig}
\usepackage{stfloats} 
\usepackage[most]{tcolorbox}
\newtcolorbox{promptbox}[1][]{
  breakable,
  enhanced,
  colback=gray!10,
  colframe=black,
  boxrule=0.5pt,
  arc=2pt,
  left=6pt,
  right=6pt,
  top=6pt,
  bottom=6pt,
  fontupper=\ttfamily\small,
  title=#1
}

\usepackage{xspace}
\newcommand{\framework}{\texttt{DPForge}\xspace}
\newcommand{\benchmark}{\texttt{DPDisc}\xspace}
\AtBeginDocument{%
  }

\renewcommand\footnotetextcopyrightpermission[1]{} 
\begin{document}

\title{DPDisc: From Factoid Questions to Data Product Requests for Open-World Data Product Discovery over Tables and Text}

\author{Liangliang Zhang$^1$, Nandana Mihindukulasooriya$^2$, Niharika D’Souza$^2$, Sola Shirai$^2$, Sarthak Dash$^2$,  Yao Ma$^1$, Horst Samulowitz$^2$}
\affiliation{
  \institution{$^1$Rensselaer Polytechnic Institute, $^2$IBM Research}
  \country{USA}
}
\email{{zhangl41, may13}@rpi.edu, {nandana, Niharika.DSouza, solashirai, sdash, samulowitz}@ibm.com}

\begin{abstract}
Data products are reusable, self-contained assets designed for specific business use cases. Automating their discovery is of great industry interest, as it enables efficient data access in large data lakes and supports analytical workflows. However, no benchmark currently exists for data product discovery over hybrid table-text corpora. Existing datasets focus on answering single factoid questions over individual tables rather than assembling multiple related data assets into coherent products. To address this gap, we present \benchmark, the first large-scale benchmark for data product discovery, where systems must retrieve coherent collections of tables and passages to satisfy high-level Data Product Requests (DPRs).

We introduce \framework, an automated pipeline that systematically repurposes table-text QA datasets by clustering related tables and passages into coherent data products, generating professional-level analytical requests using an LLM ensemble, and validating quality through multi-phase LLM evaluation. \benchmark comprises 13,076 validated instances with full provenance, derived from three representative datasets spanning open-domain and financial domains. Baseline experiments with sparse, dense, and hybrid retrieval methods imply evaluation feasibility while revealing substantial performance gaps across domains, indicating opportunities for future research in structure-aware data product discovery.
\\
\textbf{Code and datasets are available at:}\\
Dataset: \url{https://huggingface.co/datasets/ibm-research/data-product-benchmark}\\
Code: \url{https://github.com/ibm/data-product-benchmark}
\end{abstract}

\keywords{Data Product Benchmark, Table-text Copora, Data Discovery}


\maketitle
\section{Introduction}
\begin{figure*}[t]
    \centering
    \includegraphics[width=0.95\textwidth]{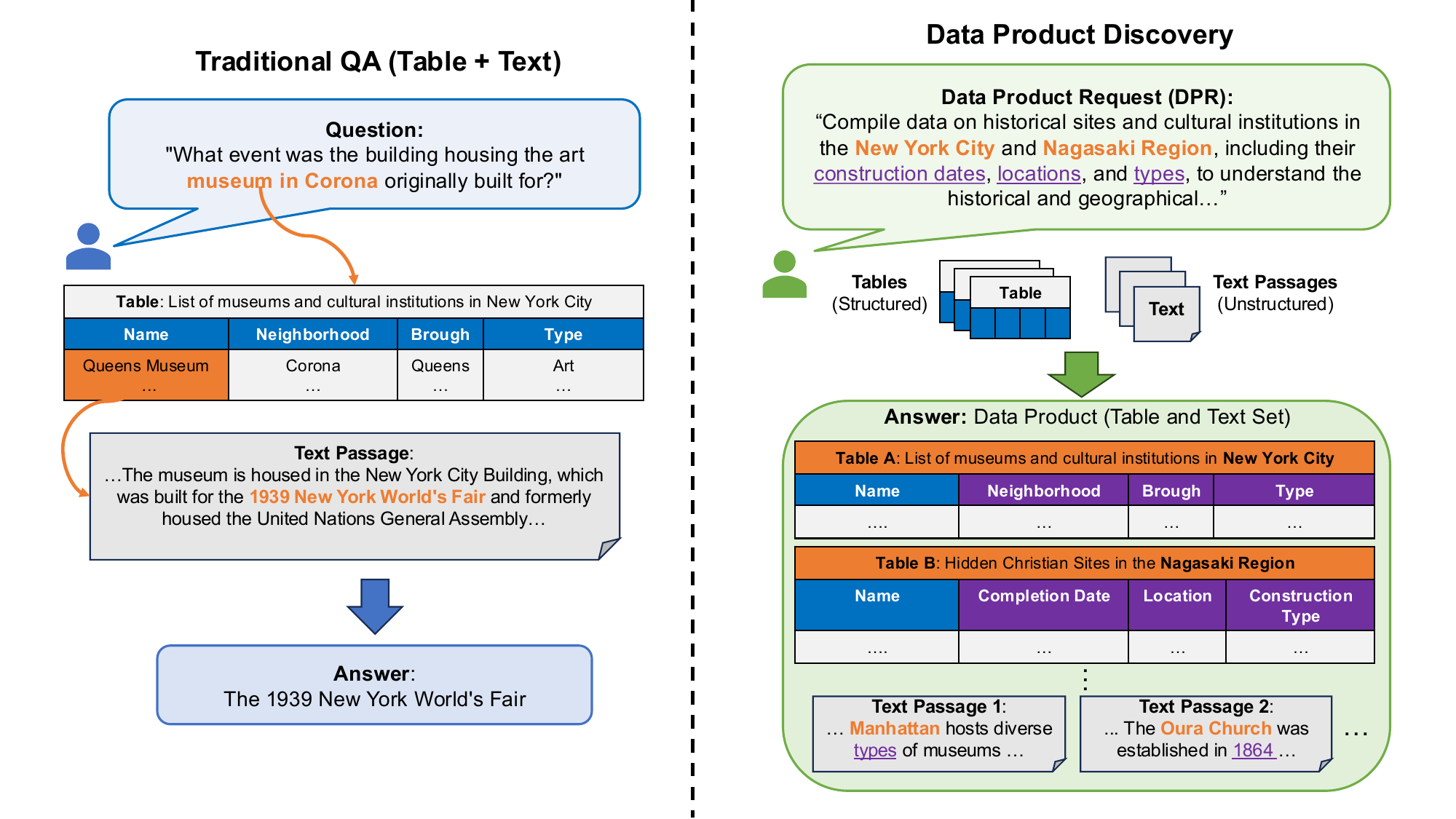}
    \caption{
    (Left) Traditional Table-Text QA tasks (e.g., HybridQA) focus on answering a factoid question from a single table with linked text. 
    (Right) \textsc{Data Product Discovery} focuses on retrieving a coherent \textit{Data Product} -- a collection of heterogeneous tables and texts. The model must align abstract user intents (e.g., data on historical/cultural sites) across diverse schemas (e.g., combining Nagasaki historical sites with NYC museums).}
    \label{fig:intro} 
\end{figure*}
Modern enterprises increasingly rely on data lakes~\cite{bogatu2020dataset} to store large volumes of heterogeneous data, ranging from structured relational tables to semi-structured and unstructured text~\cite{giebler2019leveraging,ravat2019data,fang2015managing}. While data lakes provide flexibility and scalability, the absence of a unifying schema, together with the rapid growth of data assets, makes it increasingly difficult to discover, understand, and utilize the right data for a given analytical task~\cite{datascience_book}. As a result, a substantial portion of analysts’ time is often spent not on analysis itself, but on locating, assessing, and assembling relevant data~\cite{chapman2020dataset}.

In many real-world scenarios, data discovery is driven by high-level analytical or business needs rather than low-level technical specifications. For example, an analyst might express a request such as: \emph{``I need data that can help me analyze customer retention patterns across different market segments over time.''} Fulfilling such a request typically requires identifying and combining multiple related data assets, including transactional records, customer profiles, segmentation metadata, and explanatory documentation. However, existing data discovery tools, such as data catalogs and semantic search systems, primarily operate at the level of individual tables or documents~\cite{bast2016semantic,mariscal2010survey,konan2024automating}. While effective for keyword-based lookup or schema filtering, these approaches struggle to bridge the semantic gap between abstract business requirements and the concrete data assets needed to satisfy them.

To address this disconnect, the notion of a \textbf{data product} has gained traction in both industry and academia~\cite{machado2022data,konan2024automating,goedegebuure2024data}. As shown in Figure~\ref{fig:intro}, a data product is a reusable, discoverable collection of data assets designed to support a specific analytical or decision-making use case. Unlike traditional question answering (QA)~\cite{allam2012question}, which typically focuses on extracting a single factoid answer from a specific table with text passages, data product discovery entails assembling a coherent set of heterogeneous resources. For instance, satisfying a request for ``data on historical and cultural sites'' requires aggregating diverse schemas—combining Nagasaki sites with NYC museums—into a unified package. However, automating this process remains difficult due to the scale and heterogeneity of modern data lakes. We term this challenge the semantic discovery gap: the fundamental mismatch between abstract user intent and low-level retrieval mechanisms.

From a research perspective, this problem is at the intersection of data discovery~\cite{raptis2019data}, information retrieval~\cite{singhal2001modern}, and multimodal reasoning over heterogeneous data sources, including tables and text~\cite{shraga2020web}.
Prior benchmarks have studied complex QA on structured and unstructured sources~\cite{WikiTableQuestions,chen2020hybridqa,nan2022fetaqa,pramanick2024spiqa}, showing the feasibility of reasoning across tables and text. However, these benchmarks are designed to evaluate the accuracy of the response, not the ability to retrieve and assemble the underlying data artifacts and interprete the user's request to the retrieval task for the data product wrap up and send back to user. In contrast, while the Data Lakehouse~\cite{harby2022data,machado2022data,armbrust2021lakehouse} provides the necessary infrastructure to store and govern vast quantities of heterogeneous data, it does not inherently solve the discovery challenge. A Lakehouse organizes data physically, but not semantically; it lacks the native capability to map high-level user requests to coherent sets of heterogeneous data assets. This calls for dedicated efforts to estabilish a a retrieval-centric benchmark with explicit data-product ground truth, so that we can systematically evaluate and compare data discovery methods. 

 However, constructing one from scratch presents substantial hurdles. First, real-world data lakes involve heterogeneous modalities (tables and text) where the boundaries of a coherent ``data product'' are often ambiguous. Second, authentic \textbf{ Data Product Requests (DPRs)} capture high-level analytical intent rather than explicit retrieval instructions, requiring complex reasoning to map abstract goals to concrete assets. Consequently, manual annotation is prohibitively expensive; asking experts to identify semantically related assets across vast corpora and validate their sufficiency is unscalable, making direct construction impractical.

To bridge this gap, we propose a systematic methodology that repurposes existing table-text QA datasets to construct benchmarks for the data product discovery task. These datasets—such as HybridQA~\cite{chen2020hybridqa}—already contain rich, human-curated alignments between natural language queries, focal tables, and supporting text passages. Our key insight is to reinterpret these granular resources at the level of data products. By clustering semantically related QA pairs, we synthesize coherent collections of tables and passages that simulate realistic data products. We then employ Large Language Models (LLMs) to generate~\cite{wang2023self, yu2023metamath, li2023starcoder, xu2023wizardlm} high-level DPRs that abstract away from individual questions, treating the clustered assets as the ground truth. This approach effectively transforms existing QA resources into a large-scale, principled benchmark for data product discovery, avoiding the prohibitive costs of manual annotation.

\noindent\textbf{Contributions.} Our work makes three key contributions: 
(1) We propose \textbf{\framework}, an automated pipeline that repurposes table-text QA datasets into data product discovery benchmarks through clustering, refinement, LLM-based generation, and multi-phase validation.
(2) We construct \textbf{\benchmark}, the first large-scale benchmark for data product discovery over hybrid table-text corpora, comprising 13,076 validated instances with full provenance to source data.
(3) We establish baseline retrieval methods using sparse, dense, and hybrid approaches, demonstrating the feasibility of evaluation while revealing current limitations and opportunities for future research.

\section{Related Work}
\label{sec:related}

\subsection{Table-Text Question Answering}
Table–text question-answering datasets study how models answer natural language questions by jointly using structured tables and unstructured text. Early work builds on text-only QA benchmarks such as SQuAD~\cite{rajpurkar2016squad} and TriviaQA~\cite{joshi2017triviaqa}, and later datasets explicitly incorporate tables as part of the evidence. HybridQA~\cite{chen2020hybridqa} introduced table-anchored questions that require reasoning over linked Wikipedia passages, while FeTaQA~\cite{nan2022fetaqa} allows free-form answers grounded in tabular content for more challenging table questions. Financial QA benchmarks, including FinQA~\cite{chen2021finqa}, ConvFinQA~\cite{chen2022convfinqa}, and TAT-QA~\cite{zhu2021tat}, further combine tables with textual narratives such as earnings reports and emphasize numerical reasoning. Related efforts also extend table–text QA to scanned documents and forms~\cite{ding2023vqa,raja2023icdar}, as well as to broader domains such as scientific literature~\cite{dasigi2021dataset, pramanick2024spiqa} and multi-document reasoning~\cite{zhu2022towards, hui2024uda, choi2025finder}. Text-to-SQL benchmarks like Spider~\cite{lei2024spider} evaluate semantic parsing over known database schemas, complementing table–text QA.

These datasets operate in a \textit{closed-world setting}, explicitly providing relevant data to answer a concrete question. Even datasets that allow multi-table retrieval (e.g., OTT-QA~\cite{chen2020open}) ultimately evaluate answer correctness rather than the completeness of discovered data artifacts. In our work, we focus on the challenge of discovering relevant data assets for high-level analytical requests.

\subsection{Data Discovery and Dataset Search}
Data Discovery involves identifying, curating, and organizing data assets to support analytical tasks, especially in enterprise environments with large, heterogeneous data lakes~\cite{ravat2019data, halevy2016goods,paton2023dataset}. Traditional systems mainly rely on metadata catalogs and keyword search, which struggle to bridge the semantic gap between high-level user intent and low-level data artifacts. Modern enterprise data discovery platforms, including Amundsen~\cite{choudhary2019amundsen} and DataHub~\cite{datahub_2025}, extend this paradigm by integrating metadata management, lineage tracking, and user feedback to support scalable exploration. Recent research further enhances these systems through automated profiling based on schemas, statistics, and data distributions, semantic representations of metadata and content~\cite{eichler2021modeling}, and interaction-driven signals that capture organizational knowledge~\cite{nargesian2018data}. 

Dataset search addresses similar challenges in broader scientific and open-data settings. Platforms such as Google Dataset Search~\cite{brickley2019google}, DataCite~\cite{brase2009datacite}, and Zenodo~\cite{sicilia2017community} enable keyword-based discovery across large repositories using structured metadata. Analyses of dataset search queries show that users often express abstract analytical goals rather than precise dataset identifiers, motivating semantic matching techniques that align queries with datasets using embeddings. Benchmarks such as TableUnion~\cite{hu2023automatic} study unionable table discovery, and open-domain table retrieval tasks~\cite{zhu2021tat, chen2020hybridqa} evaluate finding relevant tables from large corpora with tabular representation learning.
More recently, researchers have begun to explore discovery scenarios that involve multi-step reasoning or interaction, such as DiscoveryBench~\cite{majumder2024discoverybench}.  
LLMs have also been applied to data discovery, for example by generating structured queries or improving retrieval through learned representations~\cite{guo2025birdie}, with recent surveys outlining broader opportunities and open challenges for LLM-based discovery systems~\cite{freire2025large}. 

In parallel, the concept of a data product has emerged as a central abstraction in modern data architectures, particularly within the data mesh paradigm~\cite{goedegebuure2024data, machado2022data}. Data products emphasize domain ownership, reusability, and discoverability, and often bundle raw data with metadata, documentation, and quality guarantees~\cite{armbrust2021lakehouse}. Unlike individual datasets or table augmentation outputs, data products encapsulate curated collections of related data assets together with contextual metadata, documentation, and quality guarantees, enabling reuse and interpretation for downstream analytical workflows.
Given the complexity of the task, existing benchmarks related to data products remain limited. DP-Bench~\cite{chowdhury2025dp} is a notable example that studies data product creation from natural-language requests over predefined enterprise schemas. Unlike our work, it focuses on constructing curated tables and derived columns, \textit{assuming relevant data sources are known apriori}, which is not generally the case.

\subsection{Information Retrieval}
Traditional retrieval methods such as BM25~\cite{robertson2009probabilistic} and TF-IDF~\cite{sparck1972statistical} provide efficient retrieval but struggle with vocabulary mismatch and semantic similarity. The advent of pretrained language models such as BERT~\cite{bert_2019} enabled contextual embeddings that capture semantic meaning beyond surface-level lexical overlap. Dual-encoder architectures~\cite{karpukhin2020dense} further improved scalability by independently encoding queries and documents into dense vectors, enabling efficient similarity search through maximum inner product search (MIPS). Benchmarks such as MS MARCO~\cite{nguyen2016msmarco} and Natural Questions~\cite{kwiatkowski2019natural} have driven progress in neural retrieval methods. Hybrid retrieval methods~\cite{luan2021sparse} combine lexical signals with semantic embeddings through score fusion techniques, balancing precision on exact matches with recall on semantically similar content. 

To improve retrieval quality, re-ranking models refine an initial set of candidates retrieved by a first-stage system. Cross-encoder architectures jointly encode query-document pairs, enabling fine-grained relevance scoring at the cost of computational efficiency~\cite{nogueira2019passage}. Multi-vector models~\cite{gillick2019learning} such as ColBERT~\cite{khattab2020colbert} offer a middle ground through late interaction, providing contextualized scoring while maintaining efficiency through precomputed document representations. These methods have shown consistent improvements over single-stage retrieval across various benchmarks. Unlike QA-driven retrieval, our setting evaluates retrieval as an end in itself, where success is measured by coverage and coherence of retrieved assets rather than answer accuracy

\subsection{LLM-based Benchmarking and Evaluation}
The use of large language models as automated evaluators, often called LLM-as-a-judge~\cite{zheng2023judging}, has emerged as a scalable alternative to human annotation for assessing quality, relevance, and correctness in tasks where traditional metrics prove insufficient. Ensemble-based approaches~\cite{gu2024survey, li2024llms} employ multiple LLMs as independent judges, aggregating decisions through majority voting to mitigate individual model biases and improve robustness. Recent benchmarks for agentic and tool-using systems, such as AgentBench~\cite{liu2023agentbench}, ToolBench~\cite{wangtoolbench}, and GAIA~\cite{mialon2023gaia}, evaluate multi-step reasoning and interaction in open-ended tasks, but focus on action selection or problem solving rather than assessing the quality and completeness of discovered data assets. As a result, they do not directly address the evaluation of data discovery or data product construction from high-level analytical requests.

Beyond evaluation, LLMs are increasingly used for benchmark \emph{construction} through synthetic data generation~\cite{wang2023self, yu2023metamath, li2023starcoder, xu2023wizardlm}. To ensure quality, multi-stage validation pipelines combine automated LLM-based filtering with selective human verification~\cite{wang2023self}. Common strategies include: (i) using multiple LLMs and retaining only high-agreement instances~\cite{gu2024survey}; (ii) applying heuristics to detect degenerate cases; and (iii) conducting human audits on random samples or high-disagreement cases.
Known limitations include position bias, verbosity bias, and self-enhancement bias~\cite{wang2024large, zheng2023judging}.

Our benchmark employs this paradigm in two phases: generating diverse data product requests using five LLMs to maximize linguistic diversity (Section~\ref{step3}), and validating quality through multi-model evaluation with majority voting (Section~\ref{step4}). By combining ensemble generation with structured validation, we leverage scalability while maintaining quality control through multi-model agreement and selective human verification.

\section{\framework: Transforming table-text QA to Data Product Discovery Instances}
\label{sec:framework}

\begin{figure*}[!t]
\centering
\includegraphics[width=0.95\textwidth]{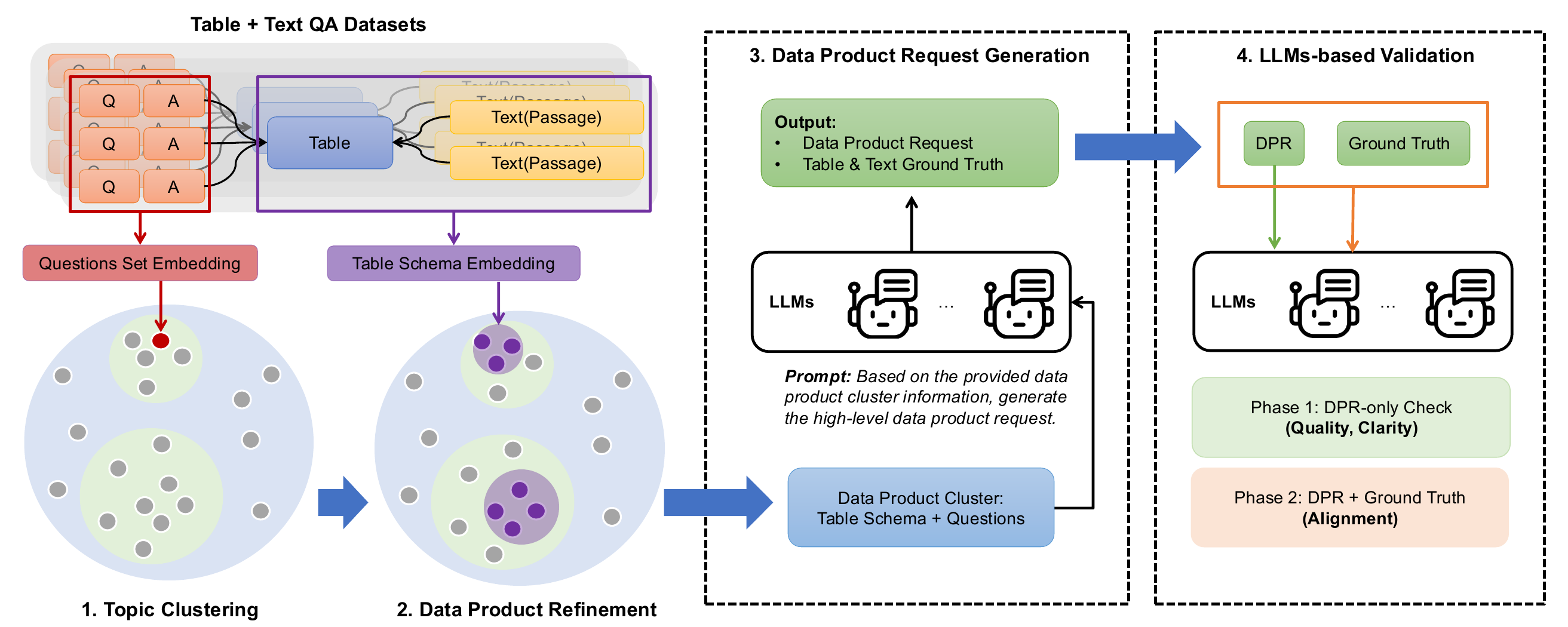}
\caption{The \framework pipeline for constructing \benchmark. We repurpose existing QA datasets via (1) topic clustering of questions, (2) schema-based data product refinement, (3) DPR generation using ensemble of LLMs, and (4) two-phase automated validation.}
\label{fig:framework}
\end{figure*}

Real-world corpora contain heterogeneous modalities—relational tables and free-form text—that must be semantically aligned at scale, yet data product boundaries remain ambiguous without explicit user intent. DPRs are inherently complex, requiring integration of multiple related tables with supporting text passages to satisfy multi-faceted analytical goals. For example, analyzing supply chain disruptions demands assembling vendor performance tables, shipment tracking records, inventory levels, and news reports—a coherent collection of interconnected assets rather than a single table or document.

While constructing DPR benchmarks directly from enterprise data lakes would be ideal, it is prohibitively expensive and labor-intensive. Existing table-text QA datasets offer a scalable alternative: they provide high-quality alignments between questions, tables, and passages across diverse domains. However, these datasets operate at narrower granularity, answering single questions about individual tables, rather than spanning multiple tables and passages as DPRs require. Our key insight is to systematically repurpose these datasets by clustering related QA instances into coherent multi-table collections and abstracting their collective intent into professional-level analytical requests.

To achieve this goal, we introduce \framework, a four-stage pipeline designed to systematically repurpose table-text QA datasets into high-fidelity data product discovery benchmarks. As illustrated in Figure~\ref{fig:framework}, the workflow proceeds through four distinct phases. First, \textbf{Topic Clustering} aggregates tables based on the semantic embeddings of their associated question sets to uncover latent domain-specific clusters. Second, \textbf{Data Product Refinement} leverages table schema embeddings to decompose coarse-grained or heterogeneous groups into semantically coherent and structurally bounded data products. Third, \textbf{Data Product Request Generation} utilizes an LLM ensemble to synthesize high-level requests that abstract the collective analytical intent of each cluster, effectively bridging the gap between granular QA pairs and holistic data product requirements. Finally, \textbf{LLMs-based Validation} implements a rigorous two-phase protocol to evaluate generated requests for intrinsic quality and clarity before verifying their factual alignment with the ground-truth tables and passages. Next, we present the details of these components in the following subsections.
\subsection{Setup and Notation}

\textbf{From QA to DPR.} 
Existing table-text QA focuses on narrow, specific questions (e.g., ``Which player scored the most goals?''). In contrast, DPRs are broader and span multiple related tables and passages (e.g., ``Collect data to analyze team performance trends over five seasons''). Bridging this gap requires moving beyond isolated QA pairs toward coherent analytical
units. To this end, we cluster semantically related table--text QA instances into data product
candidates and generate realistic DPRs that abstract the underlying analytical intent
shared by the cluster.

\noindent\textbf{Table-Text QA Datasets.} 
We build upon existing table--text QA datasets as the foundation of our benchmark.
Let $\mathcal{T}$ denote the set of tables, $\mathcal{Q}$ the set of questions, and
$\mathcal{X}$ the set of textual passages.
Each question $q \in \mathcal{Q}$ and passage $x \in \mathcal{X}$ is anchored to exactly
one table $t \in \mathcal{T}$, as illustrated in Figure~\ref{fig:framework}.
For a given table $t$, we denote the associated questions and passages by
$\mathcal{Q}(t) \subseteq \mathcal{Q}$ and $\mathcal{X}(t) \subseteq \mathcal{X}$,
respectively.

\subsection{Topic Clustering}
\label{step1}

Our goal is to transform isolated table-text QA instances into coherent data products that reflect realistic analytical scenarios. A key challenge is that existing QA datasets organize questions around individual tables scattered across heterogeneous, often unrelated topics, making it difficult to identify which tables should be grouped together. We observe that questions explicitly encode user information needs: tables attracting semantically similar questions likely belong to the same analytical domain and can form coherent data products. For example, tables associated with questions about ``player performance'', ``team rankings'', and ``season statistics'' collectively support sports analytics. This question-driven perspective enables us to infer latent topical structure without relying on table schemas or manual annotations, providing a scalable foundation for clustering related tables into meaningful analytical units.
For each table $t \in \mathcal{T}$ with at least one question, we construct a document by concatenating its questions, as $d_t = \mathrm{concat}(\mathcal{Q}(t))$.

Let $\mathcal{D} = \{d_t : t \in \mathcal{T}, \mathcal{Q}(t) \neq \varnothing\}$ denote the collection of such documents, and let $f_{\text{emb}}(\cdot)$ be a sentence embedding function. We apply \textsc{BERTopic}~\cite{grootendorst2022bertopic} to cluster the embedded documents $\{f_{\text{emb}}(d_t) : d_t \in \mathcal{D}\}$ into $K_{\text{raw}}$ semantic topics:
\begin{equation}
\phi_{\text{topic}} : \mathcal{D} \to \{1, \ldots, K_{\text{raw}}\}
\end{equation}

This yields raw clusters of tables grouped by question similarity:
\begin{equation}
\mathcal{C}^{\text{raw}} = \{C_k\}_{k=1}^{K_{\text{raw}}}, \quad 
C_k = \{t \in \mathcal{T} : \phi_{\text{topic}}(d_t) = k\}
\end{equation}

These clusters provide an initial question-driven organization of the corpus, which we later refine into coherent data products.

\subsection{Data Product Refinement}
\label{step2}

The raw topic clusters $\mathcal{C}^{\mathrm{raw}}$ from
Section~\ref{step1} are derived solely from question semantics and are therefore often coarse-grained.
A single cluster may contain heterogeneous tables or be too large to form a usable data product. For example, a topic labeled ``human'' may mix tables about athletes, historical figures, and medical patients, which differ substantially in schema and analytical use.

To construct coherent and well-scoped data products, we refine raw clusters using table
schema information and explicit size control.
The key intuition is that tables within the same topic and with similar column
structures are more likely to support a shared analytical workflow.
Let $s(t)$ denote the schema of table $t$, represented by its column headers, and let
$\mathbf{e}_t = f_{\mathrm{emb}}(s(t))$ be its schema embedding.
For a raw cluster $C_k$ of size $n_k$, we retain it as a single product if $n_k \leq K$,
where $K$ is the maximum number of tables per product; otherwise, we partition $C_k$
into $M_k = \lceil n_k / K \rceil$ subclusters using $k$-means over
$\{\mathbf{e}_t \mid t \in C_k\}$.

This process yields a set of refined data products
$\mathcal{P} = \{ \mathcal{P}_i \}_{i=1}^{K_{\mathrm{prod}}}$, where each
$\mathcal{P}_i \subseteq \mathcal{T}$ consists of semantically and structurally coherent
tables.
The supporting passages for each product are defined as
\begin{equation}
\mathcal{X}(\mathcal{P}_i) = \bigcup_{t \in \mathcal{P}_i} \mathcal{X}(t),    
\end{equation}
which serve as ground-truth context for downstream evaluation.

\subsection{Data Product Request Generation}
\label{step3}

Given a refined data product $\mathcal{P}_i$, we aim to generate a professional DPR that captures the high-level analytical intent supported by its constituent tables and passages. Unlike factoid-style questions, DPRs should describe realistic analysis tasks requiring joint use of multiple tables and contextual text. To inform generation, we construct a compact summary $\pi(\mathcal{P}_i)$ combining table schemas (column headers) and their associated questions—two complementary signals that together encode both available structured variables and typical query patterns. We employ an ensemble of LLMs to translate $\pi(\mathcal{P}_i)$ into diverse natural-language DPR candidates, as relying on a single model or fixed templates yields repetitive phrasing and limited linguistic diversity. Sampling from multiple models with different training data and inductive biases produces requests that vary in structure, terminology, and abstraction level, more closely reflecting how real analysts articulate information needs~\cite{majumder2024discoverybench}. This ensemble approach also mitigates generator-specific biases, reducing the risk that benchmark performance reflects overfitting to a particular model's idiosyncrasies rather than genuine data product discovery capability.

Specifically, we used Llama~3.3~70B~\cite{grattafiori2024llama},
DeepSeek~V3~\cite{guo2025deepseek}, GPT OSS~120B~\cite{achiam2023gpt}, Qwen~2.5~72B~\cite{yang2025qwen3}, and Mistral~8$\times$22B~\cite{jiang2023mistral7b}.
Let $\mathcal{M}_{\mathrm{gen}} = \{ M_j \}_{j=1}^{G}$ denote the set of generation models.
For each data product $\mathcal{P}_i$, we generate one candidate request per model:
\begin{equation}
r_{i,j} = \mathcal{G}_{M_j}\!\big( \pi(\mathcal{P}_i) \big), \quad j = 1, \ldots, G.
\end{equation}
The resulting candidate set is denoted by
\[
\mathcal{R}_i = \{ r_{i,1}, \ldots, r_{i,G} \}.
\]

Each generated request is explicitly grounded by pairing it with the underlying data
product and its supporting passages, yielding the ground-truth context
\[
\mathrm{GT}_i = \big( \mathcal{P}_i,\ \mathcal{X}(\mathcal{P}_i) \big).
\]
This design ensures end-to-end traceability from every DPR to concrete tables and textual
evidence, which is critical for reliable benchmark evaluation.
The full generation prompt is provided in Appendix~\ref{app:prompts}.

\subsection{LLMs-based Validation }
\label{step4}
The DPR generation step produces multiple requests per data product, which inevitably vary in quality, specificity, and alignment with the underlying data. To ensure retained DPRs are well-formed, analytically meaningful, and correctly grounded, we employ a structured validation procedure using LLMs as automated judges~\cite{gu2024survey}. We use the same ensemble of five LLMs from Section~\ref{step3} as independent evaluators, each assessing DPRs without access to other judges' outputs. Final decisions are made by majority voting across models, mitigating individual model biases and ensuring robust evaluation. Validation proceeds in two sequential phases.

\noindent\textbf{Phase 1: DPR-only screening.}
Each DPR is first evaluated in isolation, without access to its associated tables or passages, to assess intrinsic request quality. Judges rate DPRs along two dimensions: (i) \emph{quality}: whether the request is high-level, use-case driven, and actionable rather than trivial, vague, or overly narrow; and (ii) \emph{clarity}: whether the request is coherent, unambiguous, and easy to interpret. Each dimension uses a fixed ordinal scale. DPRs whose aggregated scores fall below predefined thresholds on either dimension are discarded, filtering out ill-posed or low-value requests before more expensive alignment checks.

\noindent \textbf{Phase 2: DPR--ground truth alignment.}
Surviving DPRs are evaluated jointly with their ground-truth tables and passages. Judges assess whether the provided data is sufficient to satisfy the request without external information. Requests exhibiting semantic mismatch, unsupported requirements, or underspecification are penalized. Judges also identify irrelevant components: when a majority agrees a table or passage does not contribute to the request, it is pruned from the ground-truth set. After score aggregation, DPRs failing alignment, quality, or clarity thresholds are removed. This two-phase validation yields a curated benchmark of well-formed, analytically meaningful DPRs tightly grounded in their corresponding data products.

This validation process yields a curated set of DPRs that are well-formed, analytically meaningful, and tightly grounded in their corresponding data products.

\section{\benchmark Benchmark}
\begin{figure}[t]
    \centering
    \includegraphics[width=0.5\textwidth]{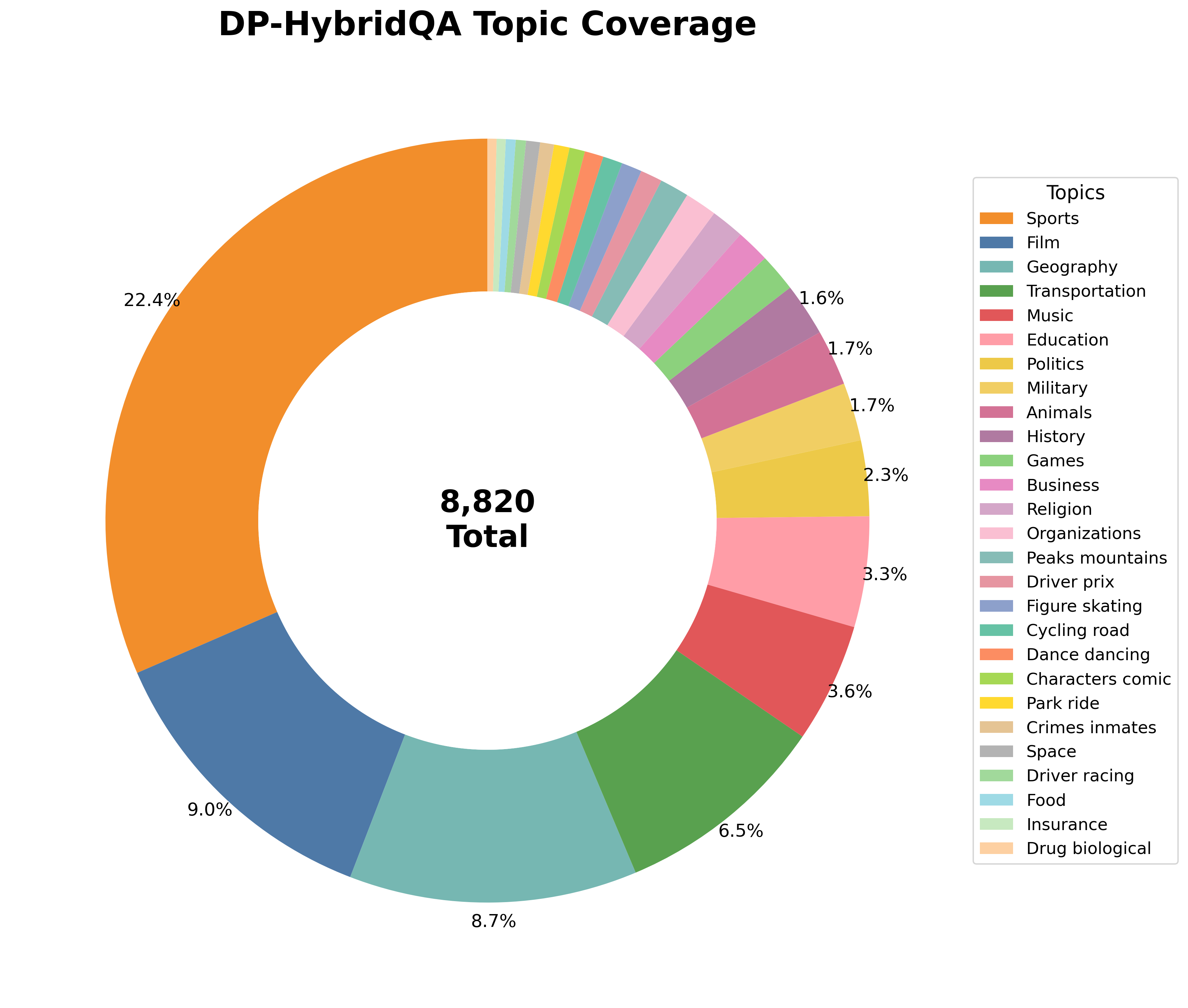}
    \caption{DP-HybridQA Topic Distribution.}
    \label{fig:hybridqa}
\end{figure}

Following the systematic transformation pipeline \framework described in Section~\ref{sec:framework}, we introduce \textbf{\benchmark} (\textbf{D}ata \textbf{P}roduct \textbf{Disc}overy Benchmark), the first large-scale benchmark specifically designed to evaluate open-world data product discovery over hybrid table-text corpora. DPDisc transforms narrow, single-table QA pairs into comprehensive data product requests that require assembling multiple related tables with supporting passages, reflecting realistic enterprise discovery scenarios.

\subsection{Dataset Characteristics}
We construct \benchmark by applying our \framework framework to three representative table-text QA datasets: the open-domain HybridQA~\cite{chen2020hybridqa} and two financial-domain datasets, TAT-QA~\cite{zhu2021tat} and ConvFinQA~\cite{chen2022convfinqa}. Through clustering, schema-based refinement, LLM-based generation, and multi-phase validation, we transform thousands of isolated QA pairs into 13,076 validated data product discovery instances. The resulting benchmark comprises three dataset variants. We name the resulting three datasets as DP-HybridQA (8,820 DPRs), DP-ConvFinQA (3,113 DPRs), and DP-TATQA (1,143 DPRs), respectively. 

\noindent{\bf Complexity and Retrieval Difficulty}. Unlike source table-text QA datasets that focus on single tables, \benchmark significantly increases retrieval difficulty by requiring multi-asset discovery. Each DPR necessitates retrieving a complete data product comprising multiple tables and passages. DP-HybridQA requests typically require 5--9 tables and 17--29 passages, reflecting the broad, interconnected nature of open-domain topics. The financial datasets are more focused but remain challenging: DP-TATQA requests require 3--4 tables and 6--7 passages on average, while DP-ConvFinQA requests require 3--7 tables and 6--12 passages. This multi-asset requirement fundamentally distinguishes data product discovery from traditional single-table retrieval.

\noindent\textbf{Topic Diversity.} To ensure \benchmark reflects real-world data lake heterogeneity, we analyzed the topical distribution of generated DPRs. DP-HybridQA exhibits broad topic coverage spanning entertainment, sports, geography, history, and science (Figure~\ref{fig:convfinqa}), mirroring the diversity of Wikipedia-sourced data. The financial datasets show distinct domain concentrations: DP-TATQA is heavily focused on corporate finance (43.9\%), while DP-ConvFinQA shows more balanced distribution between financial markets (22.5\%) and corporate finance (21.8\%) (Figure~\ref{fig:financial-topics} in Appendix). This variation enables evaluation across both broad, heterogeneous corpora and specialized, domain-focused collections.

\subsection{Human Quality Assessment}
To evaluate the quality of the DPR instances in \benchmark, we adopted a multi-faceted evaluation protocol to evaluate the DPRs and their relevance for the data-products obtained from our algorithm across different datasets. Specifically, Human annotators assessed the generated DPRs and their alignment with the corresponding data products across four key dimensions: (1) \textbf{Quality}: The level of abstraction and actionability of the DPR (binary 0/1); (2) \textbf{Clarity}: Whether the wording of the DPR is unambiguous, coherent, and implementable (Binary 0/1); (3) \textbf{Relevance}: The proportion of ground-truth tables or passages deemed relevant to the DPR (Continuous in [0,1]); and (4) \textbf{Completeness}: Whether the ground truth DP is complete or sufficient for the DPR (binary 0/1).

\noindent{\bf Annotation Design.} To promote dataset coverage while ensuring the robustness of observed trends~\cite{wang2024revisiting}, we grouped six annotators (A--F) into two independent cohorts: Group 1 (A--B--C) and Group 2 (D--E--F). This two-group structure serves as a cross-validation mechanism; by examining whether trends are consistent between independent rater pools, we verify that the quality of the \benchmark is an intrinsic property of the framework rather than an artifact of a specific group's bias. Each annotator received 51 DPRs, stratified by size and dataset. Within each group, 36 DPRs formed an ``overlapping set'' to measure inter-annotator agreement, while 15 formed a "single set" to increase total coverage. As is standard practice in~\cite{nan2022fetaqa}, we processed the 306 resulting annotations by filtering overlapping sets per pair and computing descriptive statistics—including Mean Absolute Difference (MAD) and 95\% Confidence Intervals (CI)—alongside Cohen’s kappa ($\kappa$). Mean absolute difference (MAD) quantifies inter-annotator agreement by averaging the absolute pairwise differences between annotators’ scores. Lower values indicate closer agreement and values are interpreted in the units of the annotation scale. $\kappa$ measures agreement for categorical labels while correcting for chance, where $\kappa \approx 0$ indicates chance-level agreement and higher values reflect stronger reliability (commonly interpreted as slight $\approx 0.01 -0.20$, fair $\approx 0.21-0.40$, moderate $\approx 0.41- 0.60$, substantial $\approx 0.61-0.80$ , and near-perfect $>0.80$).

\noindent{\bf Quantative Analysis.} Empirical results in Table~\ref{tab:combined_annotation} validate the benchmark's quality and reliability. (1) High Quality \& Consistency: Mean scores are high for Quality (0.88), Clarity (0.90), and Relevance (0.95). Stability across both rater cohorts—supported by tight 95\% CIs and low MAD—indicates that the generated DPRs possess an intrinsic professional quality independent of the specific rater group. (2) Agreement Trends: While $\kappa$ varied (Group D-E-F showed moderate agreement on quality/clarity vs. Group A-B-C’s slight agreement), the stability of aggregate means suggests that while analytical requests are more subjective than factoid QA, the framework consistently produces highly relevant and clear results. (3) The Completeness Challenge: Completeness yielded the lowest mean (0.64) and highest variance (0.48). This consistent lack of consensus highlights that "completeness" in open-world data discovery is a non-trivial, subjective task. Rather than reflecting data noise, this subjectivity confirms that DPDisc transcends simplistic binary QA, mirroring real-world complexity where data utility is often a matter of professional judgment.

\begin{table}[t]
\centering
\caption{\benchmark Annotation Analysis. (Top) Descriptive statistics: Mean (Std). (Bottom) Inter-annotator agreement: MAD [95\% CI] and Cohen's $\kappa$. Relevance binarized ($\ge 0.5 \to 1$).}
\label{tab:combined_annotation}
\small
\setlength{\tabcolsep}{3pt}
\begin{tabular}{l cccc}
\toprule
\textbf{Subset / Metric} & \textbf{Quality} & \textbf{Clarity} & \textbf{Relevance} & \textbf{Compl.} \\
\midrule
\rowcolor{gray!10} \multicolumn{5}{l}{\textit{Descriptive Statistics: Mean (Std)}} \\
Overall ($n=306$) & 0.88 (.33) & 0.90 (.31) & 0.95 (.19) & 0.64 (.48) \\
ConvFinQA & 0.92 (.28) & 0.89 (.32) & 0.93 (.25) & 0.74 (.44) \\
HybridQA & 0.87 (.34) & 0.90 (.30) & 0.95 (.23) & 0.61 (.49) \\
TATQA & 0.91 (.30) & 0.91 (.30) & 1.00 (.00) & 0.71 (.46) \\
\midrule
\rowcolor{gray!10} \multicolumn{5}{l}{\textit{Agreement: MAD [95\% CI] and $\kappa$}} \\
Gr. A-B-C: MAD & 0.15 \tiny{[.08, .22]} & 0.20 \tiny{[.13, .28]} & 0.04 \tiny{[.01, .07]} & 0.39 \tiny{[.30, .48]} \\
\quad \quad \quad \quad \quad $\kappa$ & 0.11 & -0.03 & 0.33 & -0.02 \\
\addlinespace[2pt]
Gr. D-E-F: MAD & 0.15 \tiny{[.08, .22]} & 0.11 \tiny{[.05, .17]} & 0.11 \tiny{[.05, .16]} & 0.48 \tiny{[.39, .58]} \\
\quad \quad \quad \quad \quad $\kappa$ & \textbf{0.48} & \textbf{0.42} & 0.09 & 0.05 \\
\bottomrule
\end{tabular}
\end{table}

\section{Establishing and Benchmarking Baselines on \benchmark }
\label{sec:experiments}

The goal of our baseline benchmarking is to establish defensible experiments for data product discovery under a controlled and reproducible protocol. We focus on evaluating how well different retrieval strategies can surface relevant tables and passages that collectively fulfill complex analytical intents. Unlike conventional document retrieval, data product discovery requires assembling multiple heterogeneous components—tables, metadata, and textual descriptions—into a coherent, task-ready unit. To ensure a fair comparison, we standardize input representations, retrieval backends, and evaluation metrics across all methods. Retrieval effectiveness is primarily assessed using recall-oriented measures at various cutoff levels, reflecting the practical need to maximize coverage.

\subsection{Retrieval Methods}
We employ a standard two-stage architecture implemented in Milvus~\cite{wang2021milvus}. To isolate metadata utility, tables are indexed solely by title and header. We evaluate three strategies: (1) \textbf{BM25}: A sparse lexical baseline ($k_1=1.5, b=0.75$); (2) \textbf{Dense Retrieval}: Captures semantic intent by encoding queries and documents with pretrained sentence transformers\footnote{\url{https://huggingface.co/sentence-transformers}}, ranking via cosine similarity; and (3) \textbf{Hybrid Retrieval}: Fuses sparse and dense ranks via Reciprocal Rank Fusion (RRF), calculated as $\text{RRF}(d) = \sum_{r \in \{r_{\text{BM25}}, r_{\text{dense}}\}} \frac{1}{k + \text{rank}_r(d)}$, where $k=60$ is a smoothing constant. This parameter-free strategy combines lexical precision with semantic generalization without requiring learned weights.

\noindent{\textbf{Re-ranking:}}
We include a neural re-ranker baseline, in which an initial list of top-$k$ candidates are re-ranked by a specialized model (BAAI/bge-reranker-v2-m3 \cite{bgereranker}). Candidate re-ranking is specific to the query, allowing the model to produce more accurate relevance scores at the cost of additional compute cost. To avoid wasteful relevance computations across the entire corpus, we use the basic retrieval strategies to collect top-500 candidatest to pass to the re-ranker model.

\noindent{\textbf{Evaluation Metrics:}}
In our setting, the goal of retrieval is not to identify a single best-matching target but to assemble a complete data product: a coherent set of tables and passages that together satisfy a higher-level analytical request. Coverage is therefore more important than precision at rank one, making recall the most appropriate metric. Therefore, we report Recall@20 and Recall@100 to simulate practical use cases where analysts inspect a small shortlist or a broader candidate set. To directly capture completeness, we also introduce Full Recall@100, which counts success only when \textit{all} required components of a data product are retrieved. The lower Full Recall reflects the challenging nature of data product discovery.

\begin{table*}[t]
\centering
\small
\setlength{\tabcolsep}{3.2pt} 
\caption{Retrieval (Top) and Reranker (Bottom) Recall Results. All experiments use granite-125m-english embeddings.}
\label{tab:results-retrieval-combined}
\begin{tabular}{l ccc ccc ccc ccc ccc}
\toprule
 & \multicolumn{6}{c}{\textbf{Tables}} & \multicolumn{6}{c}{\textbf{Text}} & \multicolumn{3}{c}{\textbf{Data Product}} \\
\cmidrule(lr){2-7}\cmidrule(lr){8-13}\cmidrule(lr){14-16}
& \multicolumn{3}{c}{Recall@20} & \multicolumn{3}{c}{Recall@100} & \multicolumn{3}{c}{Recall@20} & \multicolumn{3}{c}{Recall@100} & \multicolumn{3}{c}{Full R@100} \\
\cmidrule(lr){2-4}\cmidrule(lr){5-7}\cmidrule(lr){8-10}\cmidrule(lr){11-13}\cmidrule(lr){14-16}
\textbf{Dataset} & BM25 & Den. & Hyb. & BM25 & Den. & Hyb. & BM25 & Den. & Hyb. & BM25 & Den. & Hyb. & BM25 & Den. & Hyb. \\
\midrule
\rowcolor{gray!15} \multicolumn{16}{l}{\textit{Retrieval}} \\
DP-HybridQA  & 0.03 & 0.44 & 0.49 & 0.62 & 0.69 & 0.71 & 0.06 & 0.25 & 0.28 & 0.36 & 0.38 & 0.49 & 0.07 & 0.11 & 0.16 \\
DP-TATQA    & 0.02 & 0.33 & 0.47 & 0.81 & 0.59 & 0.79 & 0.04 & 0.28 & 0.42 & 0.73 & 0.61 & 0.72 & 0.38 & 0.18 & 0.34 \\
DP-ConvFinQA & 0.01 & 0.89 & 0.94 & 0.97 & 0.96 & 0.97 & 0.01 & 0.86 & 0.91 & 0.98 & 0.94 & 0.98 & 0.94 & 0.89 & 0.94 \\
\midrule
\rowcolor{gray!15} \multicolumn{16}{l}{\textit{Reranker}} \\
DP-HybridQA  & 0.49 & 0.48 & 0.48 & 0.68 & 0.69 & 0.69 & 0.31 & 0.31 & 0.31 & 0.49 & 0.51 & 0.52 & 0.17 & 0.18 & 0.20 \\
DP-TATQA    & 0.53 & 0.51 & 0.53 & 0.81 & 0.75 & 0.79 & 0.42 & 0.39 & 0.40 & 0.74 & 0.71 & 0.74 & 0.38 & 0.31 & 0.38 \\
DP-ConvFinQA & 0.94 & 0.94 & 0.94 & 0.98 & 0.98 & 0.98 & 0.90 & 0.90 & 0.90 & 0.97 & 0.97 & 0.97 & 0.94 & 0.95 & 0.94 \\
\bottomrule
\end{tabular}
\end{table*}

\begin{table*}[t]
\centering
\small
\setlength{\tabcolsep}{3.8pt}
\caption{Impact of Embedding Models on DP-HybridQA.}
\label{tab:results-embeddings}
\begin{tabular}{lc c cc cc cc cc cc}
\toprule
 & \textbf{Model} & \textbf{Dim} & \multicolumn{4}{c}{\textbf{Tables}} & \multicolumn{4}{c}{\textbf{Text}} & \multicolumn{2}{c}{\textbf{Data Prod.}} \\
\cmidrule(lr){4-7}\cmidrule(lr){8-11}\cmidrule(lr){12-13}
\textbf{Embedding Model} & \textbf{Params} & \textbf{Size} & \multicolumn{2}{c}{R@20} & \multicolumn{2}{c}{R@100} & \multicolumn{2}{c}{R@20} & \multicolumn{2}{c}{R@100} & \multicolumn{2}{c}{Full R@100} \\
\cmidrule(lr){4-5}\cmidrule(lr){6-7}\cmidrule(lr){8-9}\cmidrule(lr){10-11}\cmidrule(lr){12-13}
 & & & Den. & Hyb. & Den. & Hyb. & Den. & Hyb. & Den. & Hyb. & Den. & Hyb. \\
\midrule
all-mpnet-base-v2~\cite{song2020mpnet} & 110M & 768 & 0.34 & 0.45 & 0.58 & 0.67 & 0.19 & 0.25 & 0.40 & 0.47 & 0.10 & 0.14 \\
granite-125m-eng~\cite{awasthy2025graniteembeddingmodels} & 125M & 768 & 0.44 & 0.49 & 0.69 & 0.71 & 0.25 & 0.28 & 0.38 & 0.49 & 0.11 & 0.16 \\
m-e5-large-inst~\cite{wang2024multilingual} & 560M & 1024 & 0.44 & 0.47 & 0.67 & 0.70 & 0.24 & 0.28 & 0.47 & 0.50 & 0.14 & 0.17 \\
Qwen3~\cite{yang2025qwen3} & 8B & 4096 & 0.43 & 0.48 & 0.66 & 0.70 & 0.26 & 0.28 & 0.48 & 0.51 & 0.17 & 0.19 \\
KaLM-Gemma3-12B~\cite{zhao2025kalmembeddingv2} & 12B & 3840 & 0.51 & 0.50 & 0.74 & 0.73 & 0.31 & 0.30 & 0.54 & 0.54 & 0.22 & 0.22 \\
\bottomrule
\end{tabular}
\end{table*}

\subsection{Results and Analysis}

\noindent\textbf{Challenge of \benchmark.} Retrieval results in Table~\ref{tab:results-retrieval-combined} (Top) show large performance variation across domains. Hybrid retrieval achieves 0.94 Full Recall@100 on \textbf{DP-ConvFinQA} but only 0.16 on \textbf{DP-HybridQA}—a 78-point gap. This difference reflects a fundamental challenge: standard retrievers work well for structured financial queries with consistent terminology, but fail when user intent is abstract and poorly aligned with table schemas. Dense retrieval outperforms BM25 on semantically complex tasks (0.44 vs. 0.03 Recall@20 on DP-HybridQA), capturing conceptual similarity beyond exact term matches. However, BM25 remains effective for entity-centric queries where lexical precision matters: it achieves 0.81 Recall@100 on \textbf{DP-TATQA}, where dates, company names, and financial terms can be matched directly. The results suggest that hybrid methods are necessary to balance semantic and lexical signals across diverse data product discovery scenarios.

\noindent\textbf{Reranking: An Equalizer, Not a Solver.} As shown in Table~\ref{tab:results-retrieval-combined} (Bottom), neural reranking serves primarily as a potent equalizer for weaker baselines. It significantly mitigates the lexical mismatch of sparse retrieval—for instance, more than doubling BM25 performance on DP-HybridQA (0.07 to 0.17 Full Recall). This indicates that relevant tables often exist in the "long tail" of lexical search results but are penalized by simple TF-IDF scoring. However, reranking fails to raise the performance ceiling of the strongest systems (e.g., DP-ConvFinQA remains saturated at 0.94). This  indicates a fundamental candidate generation bottleneck: if the relevant data product components are not retrieved in the top-500 pool during the first stage, the reranker is rendered useless. Consequently, improvements in downstream ranking cannot compensate for recall failures in the indexing stage, indicating the need for better query expansion or dense retrieval strategies.

\noindent\textbf{Scaling Laws vs. Structural Alignment.} Table~\ref{tab:results-embeddings} shows that larger embedding models provide limited gains. Scaling from 110M to 12B parameters improves Hybrid Recall@20 by only $\sim$7\%, despite much higher computational cost for storing and querying embeddings. This suggests that model size alone does not solve data product discovery. General-purpose embeddings trained on prose likely miss the structural properties of tables—such as relationships between headers and cell values, or dependencies across columns. The core challenge is not semantic capacity but the mismatch between abstract user requests and concrete table schemas. Better performance likely requires structure-aware pretraining or domain-specific adaptation rather than simply scaling model size.

\section{Conclusion}
\label{sec:conclusion}
We presented \benchmark, the first benchmark designed for data product discovery over table–text corpora. Our end-to-end pipeline \framework converts QA datasets into realistic DPR instances through table clustering, data product refinement, LLM-based request generation, and quality validation. Generated DP-HybridQA, DP-TATQA, and DP-ConvFinQA, this framework produces validated DPR-data product pairs with full provenance across open-domain and financial settings. We established baselines using sparse, dense, and hybrid retrieval methods. Experiments show the benchmark's discriminative power: standard methods achieve 0.94 Recall on structured financial queries (DP-ConvFinQA) but only 0.16 on open-domain requests (DP-HybridQA). Results with large embeddings (12B parameters) and neural reranking reveal a candidate generation bottleneck—scaling model size or improving ranking cannot fix poor initial retrieval. This points to the need for structure-aware indexing and retrieval methods designed specifically for data product discovery.

\noindent{\textbf{Limitations.}} Although \benchmark effectively transforms QA datasets into structured benchmark instances, several limitations remain. Its construction depends on the quality and coverage of the source datasets, which may limit the diversity of analytical scenarios. Current evaluation focuses mainly on retrieval accuracy, leaving out other important dimensions such as coherence, analytical usefulness, or user experience. In addition, the domain-specific variations observed in our experiments suggest that \benchmark may require domain-adaptive tuning to achieve optimal performance.

\bibliographystyle{ACM-Reference-Format}
\newpage
\bibliography{references}


\begin{thebibliography}{82}


\ifx \showCODEN    \undefined \def \showCODEN     #1{\unskip}     \fi
\ifx \showDOI      \undefined \def \showDOI       #1{#1}\fi
\ifx \showISBNx    \undefined \def \showISBNx     #1{\unskip}     \fi
\ifx \showISBNxiii \undefined \def \showISBNxiii  #1{\unskip}     \fi
\ifx \showISSN     \undefined \def \showISSN      #1{\unskip}     \fi
\ifx \showLCCN     \undefined \def \showLCCN      #1{\unskip}     \fi
\ifx \shownote     \undefined \def \shownote      #1{#1}          \fi
\ifx \showarticletitle \undefined \def \showarticletitle #1{#1}   \fi
\ifx \showURL      \undefined \def \showURL       {\relax}        \fi
\providecommand\bibfield[2]{#2}
\providecommand\bibinfo[2]{#2}
\providecommand\natexlab[1]{#1}
\providecommand\showeprint[2][]{arXiv:#2}

\bibitem[Achiam et~al\mbox{.}(2023)]%
        {achiam2023gpt}
\bibfield{author}{\bibinfo{person}{Josh Achiam}, \bibinfo{person}{Steven Adler}, \bibinfo{person}{Sandhini Agarwal}, \bibinfo{person}{Lama Ahmad}, \bibinfo{person}{Ilge Akkaya}, \bibinfo{person}{Florencia~Leoni Aleman}, \bibinfo{person}{Diogo Almeida}, \bibinfo{person}{Janko Altenschmidt}, \bibinfo{person}{Sam Altman}, \bibinfo{person}{Shyamal Anadkat}, {et~al\mbox{.}}} \bibinfo{year}{2023}\natexlab{}.
\newblock \showarticletitle{Gpt-4 technical report}.
\newblock \bibinfo{journal}{\emph{arXiv preprint arXiv:2303.08774}} (\bibinfo{year}{2023}).
\newblock


\bibitem[Allam and Haggag(2012)]%
        {allam2012question}
\bibfield{author}{\bibinfo{person}{Ali Mohamed~Nabil Allam} {and} \bibinfo{person}{Mohamed~Hassan Haggag}.} \bibinfo{year}{2012}\natexlab{}.
\newblock \showarticletitle{The question answering systems: A survey}.
\newblock \bibinfo{journal}{\emph{International Journal of Research and Reviews in Information Sciences (IJRRIS)}} \bibinfo{volume}{2}, \bibinfo{number}{3} (\bibinfo{year}{2012}).
\newblock


\bibitem[Armbrust et~al\mbox{.}(2021)]%
        {armbrust2021lakehouse}
\bibfield{author}{\bibinfo{person}{Michael Armbrust}, \bibinfo{person}{Ali Ghodsi}, \bibinfo{person}{Reynold Xin}, \bibinfo{person}{Matei Zaharia}, {et~al\mbox{.}}} \bibinfo{year}{2021}\natexlab{}.
\newblock \showarticletitle{Lakehouse: a new generation of open platforms that unify data warehousing and advanced analytics}. In \bibinfo{booktitle}{\emph{Proceedings of CIDR}}, Vol.~\bibinfo{volume}{8}. \bibinfo{pages}{28}.
\newblock


\bibitem[Bast et~al\mbox{.}(2016)]%
        {bast2016semantic}
\bibfield{author}{\bibinfo{person}{Hannah Bast}, \bibinfo{person}{Bj{\"o}rn Buchhold}, \bibinfo{person}{Elmar Haussmann}, {et~al\mbox{.}}} \bibinfo{year}{2016}\natexlab{}.
\newblock \showarticletitle{Semantic search on text and knowledge bases}.
\newblock \bibinfo{journal}{\emph{Foundations and Trends{\textregistered} in Information Retrieval}} \bibinfo{volume}{10}, \bibinfo{number}{2-3} (\bibinfo{year}{2016}), \bibinfo{pages}{119--271}.
\newblock


\bibitem[Bogatu et~al\mbox{.}(2020)]%
        {bogatu2020dataset}
\bibfield{author}{\bibinfo{person}{Alex Bogatu}, \bibinfo{person}{Alvaro~AA Fernandes}, \bibinfo{person}{Norman~W Paton}, {and} \bibinfo{person}{Nikolaos Konstantinou}.} \bibinfo{year}{2020}\natexlab{}.
\newblock \showarticletitle{Dataset discovery in data lakes}. In \bibinfo{booktitle}{\emph{2020 ieee 36th international conference on data engineering (icde)}}. IEEE, \bibinfo{pages}{709--720}.
\newblock


\bibitem[Brase(2009)]%
        {brase2009datacite}
\bibfield{author}{\bibinfo{person}{Jan Brase}.} \bibinfo{year}{2009}\natexlab{}.
\newblock \showarticletitle{DataCite-A global registration agency for research data}. In \bibinfo{booktitle}{\emph{2009 fourth international conference on cooperation and promotion of information resources in science and technology}}. IEEE, \bibinfo{pages}{257--261}.
\newblock


\bibitem[Brickley et~al\mbox{.}(2019)]%
        {brickley2019google}
\bibfield{author}{\bibinfo{person}{Dan Brickley}, \bibinfo{person}{Matthew Burgess}, {and} \bibinfo{person}{Natasha Noy}.} \bibinfo{year}{2019}\natexlab{}.
\newblock \showarticletitle{Google dataset search: Building a search engine for datasets in an open web ecosystem}. In \bibinfo{booktitle}{\emph{The world wide web conference}}. \bibinfo{pages}{1365--1375}.
\newblock


\bibitem[Chapman et~al\mbox{.}(2020)]%
        {chapman2020dataset}
\bibfield{author}{\bibinfo{person}{Adriane Chapman}, \bibinfo{person}{Elena Simperl}, \bibinfo{person}{Laura Koesten}, \bibinfo{person}{George Konstantinidis}, \bibinfo{person}{Luis-Daniel Ib{\'a}{\~n}ez}, \bibinfo{person}{Emilia Kacprzak}, {and} \bibinfo{person}{Paul Groth}.} \bibinfo{year}{2020}\natexlab{}.
\newblock \showarticletitle{Dataset search: a survey}.
\newblock \bibinfo{journal}{\emph{The VLDB Journal}} \bibinfo{volume}{29}, \bibinfo{number}{1} (\bibinfo{year}{2020}), \bibinfo{pages}{251--272}.
\newblock


\bibitem[Chen et~al\mbox{.}(2024)]%
        {bgereranker}
\bibfield{author}{\bibinfo{person}{Jianlyu Chen}, \bibinfo{person}{Shitao Xiao}, \bibinfo{person}{Peitian Zhang}, \bibinfo{person}{Kun Luo}, \bibinfo{person}{Defu Lian}, {and} \bibinfo{person}{Zheng Liu}.} \bibinfo{year}{2024}\natexlab{}.
\newblock \showarticletitle{{M}3-Embedding: Multi-Linguality, Multi-Functionality, Multi-Granularity Text Embeddings Through Self-Knowledge Distillation}. In \bibinfo{booktitle}{\emph{Findings of the Association for Computational Linguistics: ACL 2024}}, \bibfield{editor}{\bibinfo{person}{Lun-Wei Ku}, \bibinfo{person}{Andre Martins}, {and} \bibinfo{person}{Vivek Srikumar}} (Eds.). \bibinfo{publisher}{Association for Computational Linguistics}, \bibinfo{address}{Bangkok, Thailand}, \bibinfo{pages}{2318--2335}.
\newblock
\urldef\tempurl%
\url{https://doi.org/10.18653/v1/2024.findings-acl.137}
\showDOI{\tempurl}


\bibitem[Chen et~al\mbox{.}(2020a)]%
        {chen2020open}
\bibfield{author}{\bibinfo{person}{Wenhu Chen}, \bibinfo{person}{Ming-Wei Chang}, \bibinfo{person}{Eva Schlinger}, \bibinfo{person}{William Wang}, {and} \bibinfo{person}{William~W Cohen}.} \bibinfo{year}{2020}\natexlab{a}.
\newblock \showarticletitle{Open question answering over tables and text}.
\newblock \bibinfo{journal}{\emph{arXiv preprint arXiv:2010.10439}} (\bibinfo{year}{2020}).
\newblock


\bibitem[Chen et~al\mbox{.}(2020b)]%
        {chen2020hybridqa}
\bibfield{author}{\bibinfo{person}{Wenhu Chen}, \bibinfo{person}{Hanwen Zha}, \bibinfo{person}{Zhiyu Chen}, \bibinfo{person}{Wenhan Xiong}, \bibinfo{person}{Hong Wang}, {and} \bibinfo{person}{William Wang}.} \bibinfo{year}{2020}\natexlab{b}.
\newblock \showarticletitle{HybridQA: A Dataset of Multi-Hop Question Answering over Tabular and Textual Data}.
\newblock \bibinfo{journal}{\emph{Findings of EMNLP 2020}} (\bibinfo{year}{2020}).
\newblock


\bibitem[Chen et~al\mbox{.}(2021)]%
        {chen2021finqa}
\bibfield{author}{\bibinfo{person}{Zhiyu Chen}, \bibinfo{person}{Wenhu Chen}, \bibinfo{person}{Charese Smiley}, \bibinfo{person}{Sameena Shah}, \bibinfo{person}{Iana Borova}, \bibinfo{person}{Dylan Langdon}, \bibinfo{person}{Reema Moussa}, \bibinfo{person}{Matt Beane}, \bibinfo{person}{Ting-Hao Huang}, \bibinfo{person}{Bryan Routledge}, {et~al\mbox{.}}} \bibinfo{year}{2021}\natexlab{}.
\newblock \showarticletitle{Finqa: A dataset of numerical reasoning over financial data}.
\newblock \bibinfo{journal}{\emph{arXiv preprint arXiv:2109.00122}} (\bibinfo{year}{2021}).
\newblock


\bibitem[Chen et~al\mbox{.}(2022)]%
        {chen2022convfinqa}
\bibfield{author}{\bibinfo{person}{Zhiyu Chen}, \bibinfo{person}{Shiyang Li}, \bibinfo{person}{Charese Smiley}, \bibinfo{person}{Zhiqiang Ma}, \bibinfo{person}{Sameena Shah}, {and} \bibinfo{person}{William~Yang Wang}.} \bibinfo{year}{2022}\natexlab{}.
\newblock \showarticletitle{ConvFinQA: Exploring the Chain of Numerical Reasoning in Conversational Finance Question Answering}.
\newblock \bibinfo{journal}{\emph{Proceedings of EMNLP 2022}} (\bibinfo{year}{2022}).
\newblock


\bibitem[Choi et~al\mbox{.}(2025)]%
        {choi2025finder}
\bibfield{author}{\bibinfo{person}{Chanyeol Choi}, \bibinfo{person}{Jihoon Kwon}, \bibinfo{person}{Jaeseon Ha}, \bibinfo{person}{Hojun Choi}, \bibinfo{person}{Chaewoon Kim}, \bibinfo{person}{Yongjae Lee}, \bibinfo{person}{Jy-yong Sohn}, {and} \bibinfo{person}{Alejandro Lopez-Lira}.} \bibinfo{year}{2025}\natexlab{}.
\newblock \showarticletitle{Finder: Financial dataset for question answering and evaluating retrieval-augmented generation}. In \bibinfo{booktitle}{\emph{Proceedings of the 6th ACM International Conference on AI in Finance}}. \bibinfo{pages}{638--646}.
\newblock


\bibitem[Choudhary and Grover(2019)]%
        {choudhary2019amundsen}
\bibfield{author}{\bibinfo{person}{Shiva Choudhary} {and} \bibinfo{person}{Mark Grover}.} \bibinfo{year}{2019}\natexlab{}.
\newblock \bibinfo{title}{Amundsen: A Data Discovery and Metadata Engine}.
\newblock \bibinfo{howpublished}{Lyft Engineering Blog}.
\newblock
\urldef\tempurl%
\url{https://www.amundsen.io/}
\showURL{%
\tempurl}


\bibitem[Chowdhury et~al\mbox{.}(2025)]%
        {chowdhury2025dp}
\bibfield{author}{\bibinfo{person}{Faisal Chowdhury}, \bibinfo{person}{Sola Shirai}, \bibinfo{person}{Sarthak Dash}, \bibinfo{person}{Nandana Mihindukulasooriya}, {and} \bibinfo{person}{Horst Samulowitz}.} \bibinfo{year}{2025}\natexlab{}.
\newblock \showarticletitle{DP-Bench: A Benchmark for Evaluating Data Product Creation Systems}.
\newblock \bibinfo{journal}{\emph{arXiv preprint arXiv:2512.15798}} (\bibinfo{year}{2025}).
\newblock


\bibitem[Dasigi et~al\mbox{.}(2021)]%
        {dasigi2021dataset}
\bibfield{author}{\bibinfo{person}{Pradeep Dasigi}, \bibinfo{person}{Kyle Lo}, \bibinfo{person}{Iz Beltagy}, \bibinfo{person}{Arman Cohan}, \bibinfo{person}{Noah~A Smith}, {and} \bibinfo{person}{Matt Gardner}.} \bibinfo{year}{2021}\natexlab{}.
\newblock \showarticletitle{A dataset of information-seeking questions and answers anchored in research papers}.
\newblock \bibinfo{journal}{\emph{arXiv preprint arXiv:2105.03011}} (\bibinfo{year}{2021}).
\newblock


\bibitem[DataHub(2025)]%
        {datahub_2025}
\bibfield{author}{\bibinfo{person}{DataHub}.} \bibinfo{year}{2025}\natexlab{}.
\newblock \bibinfo{title}{Modern Data Catalog \& Metadata Platform}.
\newblock \bibinfo{howpublished}{\url{https://datahub.com/}}.
\newblock
\newblock
\shownote{Accessed: 2025-09-23}.


\bibitem[Devlin et~al\mbox{.}(2019)]%
        {bert_2019}
\bibfield{author}{\bibinfo{person}{Jacob Devlin}, \bibinfo{person}{Ming-Wei Chang}, \bibinfo{person}{Kenton Lee}, {and} \bibinfo{person}{Kristina Toutanova}.} \bibinfo{year}{2019}\natexlab{}.
\newblock \showarticletitle{BERT: Pre-training of Deep Bidirectional Transformers for Language Understanding}.
\newblock \bibinfo{journal}{\emph{Proceedings of the 2019 Conference of the North American Chapter of the Association for Computational Linguistics: Human Language Technologies}}  \bibinfo{volume}{1} (\bibinfo{year}{2019}), \bibinfo{pages}{4171--4186}.
\newblock
\urldef\tempurl%
\url{https://doi.org/10.18653/v1/N19-1423}
\showDOI{\tempurl}


\bibitem[Ding et~al\mbox{.}(2023)]%
        {ding2023vqa}
\bibfield{author}{\bibinfo{person}{Yihao Ding}, \bibinfo{person}{Siwen Luo}, \bibinfo{person}{Hyunsuk Chung}, {and} \bibinfo{person}{Soyeon~Caren Han}.} \bibinfo{year}{2023}\natexlab{}.
\newblock \showarticletitle{VQA: A new dataset for real-world VQA on PDF documents}. In \bibinfo{booktitle}{\emph{Joint European Conference on Machine Learning and Knowledge Discovery in Databases}}. Springer, \bibinfo{pages}{585--601}.
\newblock


\bibitem[Eichler et~al\mbox{.}(2021)]%
        {eichler2021modeling}
\bibfield{author}{\bibinfo{person}{Rebecca Eichler}, \bibinfo{person}{Corinna Giebler}, \bibinfo{person}{Christoph Gr{\"o}ger}, \bibinfo{person}{Holger Schwarz}, {and} \bibinfo{person}{Bernhard Mitschang}.} \bibinfo{year}{2021}\natexlab{}.
\newblock \showarticletitle{Modeling metadata in data lakes—A generic model}.
\newblock \bibinfo{journal}{\emph{Data \& knowledge engineering}}  \bibinfo{volume}{136} (\bibinfo{year}{2021}), \bibinfo{pages}{101931}.
\newblock


\bibitem[et~al.(2025)]%
        {awasthy2025graniteembeddingmodels}
\bibfield{author}{\bibinfo{person}{Parul~Awasthy et al.}} \bibinfo{year}{2025}\natexlab{}.
\newblock \bibinfo{title}{Granite Embedding Models}.
\newblock
\newblock
\showeprint[arxiv]{2502.20204}~[cs.IR]
\urldef\tempurl%
\url{https://arxiv.org/abs/2502.20204}
\showURL{%
\tempurl}


\bibitem[Fang(2015)]%
        {fang2015managing}
\bibfield{author}{\bibinfo{person}{Huang Fang}.} \bibinfo{year}{2015}\natexlab{}.
\newblock \showarticletitle{Managing data lakes in big data era: What's a data lake and why has it became popular in data management ecosystem}. In \bibinfo{booktitle}{\emph{2015 IEEE International Conference on Cyber Technology in Automation, Control, and Intelligent Systems (CYBER)}}. IEEE, \bibinfo{pages}{820--824}.
\newblock


\bibitem[Freire et~al\mbox{.}(2025)]%
        {freire2025large}
\bibfield{author}{\bibinfo{person}{Juliana Freire}, \bibinfo{person}{Grace Fan}, \bibinfo{person}{Benjamin Feuer}, \bibinfo{person}{Christos Koutras}, \bibinfo{person}{Yurong Liu}, \bibinfo{person}{Eduardo Pe{\~n}a}, \bibinfo{person}{A{\'e}cio~SR Santos}, \bibinfo{person}{Cl{\'a}udio~T Silva}, {and} \bibinfo{person}{Eden Wu}.} \bibinfo{year}{2025}\natexlab{}.
\newblock \showarticletitle{Large Language Models for Data Discovery and Integration: Challenges and Opportunities.}
\newblock \bibinfo{journal}{\emph{IEEE Data Eng. Bull.}} \bibinfo{volume}{49}, \bibinfo{number}{1} (\bibinfo{year}{2025}), \bibinfo{pages}{3--31}.
\newblock


\bibitem[Giebler et~al\mbox{.}(2019)]%
        {giebler2019leveraging}
\bibfield{author}{\bibinfo{person}{Corinna Giebler}, \bibinfo{person}{Christoph Gr{\"o}ger}, \bibinfo{person}{Eva Hoos}, \bibinfo{person}{Holger Schwarz}, {and} \bibinfo{person}{Bernhard Mitschang}.} \bibinfo{year}{2019}\natexlab{}.
\newblock \showarticletitle{Leveraging the data lake: current state and challenges}. In \bibinfo{booktitle}{\emph{International Conference on Big Data Analytics and Knowledge Discovery}}. Springer, \bibinfo{pages}{179--188}.
\newblock


\bibitem[Gillick et~al\mbox{.}(2019)]%
        {gillick2019learning}
\bibfield{author}{\bibinfo{person}{Dan Gillick}, \bibinfo{person}{Sayali Kulkarni}, \bibinfo{person}{Larry Lansing}, \bibinfo{person}{Alessandro Presta}, \bibinfo{person}{Jason Baldridge}, \bibinfo{person}{Eugene Ie}, {and} \bibinfo{person}{Diego Garcia-Olano}.} \bibinfo{year}{2019}\natexlab{}.
\newblock \showarticletitle{Learning dense representations for entity retrieval}. In \bibinfo{booktitle}{\emph{Proceedings of the 23rd conference on computational natural language learning (CoNLL)}}. \bibinfo{pages}{528--537}.
\newblock


\bibitem[Goedegebuure et~al\mbox{.}(2024)]%
        {goedegebuure2024data}
\bibfield{author}{\bibinfo{person}{Abel Goedegebuure}, \bibinfo{person}{Indika Kumara}, \bibinfo{person}{Stefan Driessen}, \bibinfo{person}{Willem-Jan Van Den~Heuvel}, \bibinfo{person}{Geert Monsieur}, \bibinfo{person}{Damian~Andrew Tamburri}, {and} \bibinfo{person}{Dario~Di Nucci}.} \bibinfo{year}{2024}\natexlab{}.
\newblock \showarticletitle{Data mesh: a systematic gray literature review}.
\newblock \bibinfo{journal}{\emph{Comput. Surveys}} \bibinfo{volume}{57}, \bibinfo{number}{1} (\bibinfo{year}{2024}), \bibinfo{pages}{1--36}.
\newblock


\bibitem[Grattafiori et~al\mbox{.}(2024)]%
        {grattafiori2024llama}
\bibfield{author}{\bibinfo{person}{Aaron Grattafiori}, \bibinfo{person}{Abhimanyu Dubey}, \bibinfo{person}{Abhinav Jauhri}, \bibinfo{person}{Abhinav Pandey}, \bibinfo{person}{Abhishek Kadian}, \bibinfo{person}{Ahmad Al-Dahle}, \bibinfo{person}{Aiesha Letman}, \bibinfo{person}{Akhil Mathur}, \bibinfo{person}{Alan Schelten}, \bibinfo{person}{Alex Vaughan}, {et~al\mbox{.}}} \bibinfo{year}{2024}\natexlab{}.
\newblock \showarticletitle{The llama 3 herd of models}.
\newblock \bibinfo{journal}{\emph{arXiv preprint arXiv:2407.21783}} (\bibinfo{year}{2024}).
\newblock


\bibitem[Grootendorst(2022)]%
        {grootendorst2022bertopic}
\bibfield{author}{\bibinfo{person}{Maarten Grootendorst}.} \bibinfo{year}{2022}\natexlab{}.
\newblock \showarticletitle{BERTopic: Neural topic modeling with a class-based TF-IDF procedure}.
\newblock \bibinfo{journal}{\emph{arXiv preprint arXiv:2203.05794}} (\bibinfo{year}{2022}).
\newblock


\bibitem[Gu et~al\mbox{.}(2024)]%
        {gu2024survey}
\bibfield{author}{\bibinfo{person}{Jiawei Gu}, \bibinfo{person}{Xuhui Jiang}, \bibinfo{person}{Zhichao Shi}, \bibinfo{person}{Hexiang Tan}, \bibinfo{person}{Xuehao Zhai}, \bibinfo{person}{Chengjin Xu}, \bibinfo{person}{Wei Li}, \bibinfo{person}{Yinghan Shen}, \bibinfo{person}{Shengjie Ma}, \bibinfo{person}{Honghao Liu}, {et~al\mbox{.}}} \bibinfo{year}{2024}\natexlab{}.
\newblock \showarticletitle{A survey on llm-as-a-judge}.
\newblock \bibinfo{journal}{\emph{arXiv preprint arXiv:2411.15594}} (\bibinfo{year}{2024}).
\newblock


\bibitem[Guo et~al\mbox{.}(2025b)]%
        {guo2025deepseek}
\bibfield{author}{\bibinfo{person}{Daya Guo}, \bibinfo{person}{Dejian Yang}, \bibinfo{person}{Haowei Zhang}, \bibinfo{person}{Junxiao Song}, \bibinfo{person}{Ruoyu Zhang}, \bibinfo{person}{Runxin Xu}, \bibinfo{person}{Qihao Zhu}, \bibinfo{person}{Shirong Ma}, \bibinfo{person}{Peiyi Wang}, \bibinfo{person}{Xiao Bi}, {et~al\mbox{.}}} \bibinfo{year}{2025}\natexlab{b}.
\newblock \showarticletitle{Deepseek-r1: Incentivizing reasoning capability in llms via reinforcement learning}.
\newblock \bibinfo{journal}{\emph{arXiv preprint arXiv:2501.12948}} (\bibinfo{year}{2025}).
\newblock


\bibitem[Guo et~al\mbox{.}(2025a)]%
        {guo2025birdie}
\bibfield{author}{\bibinfo{person}{Yuxiang Guo}, \bibinfo{person}{Zhonghao Hu}, \bibinfo{person}{Yuren Mao}, \bibinfo{person}{Baihua Zheng}, \bibinfo{person}{Yunjun Gao}, {and} \bibinfo{person}{Mingwei Zhou}.} \bibinfo{year}{2025}\natexlab{a}.
\newblock \showarticletitle{Birdie: Natural Language-Driven Table Discovery Using Differentiable Search Index}.
\newblock \bibinfo{journal}{\emph{arXiv preprint arXiv:2504.21282}} (\bibinfo{year}{2025}).
\newblock


\bibitem[Halevy et~al\mbox{.}(2016)]%
        {halevy2016goods}
\bibfield{author}{\bibinfo{person}{Alon Halevy}, \bibinfo{person}{Flip Korn}, \bibinfo{person}{Natasha~F. Noy}, \bibinfo{person}{Christopher Olston}, \bibinfo{person}{Neoklis Polyzotis}, \bibinfo{person}{Sudip Roy}, {and} \bibinfo{person}{Steven~Euijong Whang}.} \bibinfo{year}{2016}\natexlab{}.
\newblock \showarticletitle{GOODS: Organizing Google’s Datasets}. In \bibinfo{booktitle}{\emph{Proceedings of the 2016 International Conference on Management of Data (SIGMOD)}}. \bibinfo{publisher}{ACM}, \bibinfo{pages}{795--806}.
\newblock


\bibitem[Harby and Zulkernine(2022)]%
        {harby2022data}
\bibfield{author}{\bibinfo{person}{Ahmed~A Harby} {and} \bibinfo{person}{Farhana Zulkernine}.} \bibinfo{year}{2022}\natexlab{}.
\newblock \showarticletitle{From data warehouse to lakehouse: A comparative review}. In \bibinfo{booktitle}{\emph{2022 IEEE international conference on big data (big data)}}. IEEE, \bibinfo{pages}{389--395}.
\newblock


\bibitem[Hu et~al\mbox{.}(2023)]%
        {hu2023automatic}
\bibfield{author}{\bibinfo{person}{Xuming Hu}, \bibinfo{person}{Shen Wang}, \bibinfo{person}{Xiao Qin}, \bibinfo{person}{Chuan Lei}, \bibinfo{person}{Zhengyuan Shen}, \bibinfo{person}{Christos Faloutsos}, \bibinfo{person}{Asterios Katsifodimos}, \bibinfo{person}{George Karypis}, \bibinfo{person}{Lijie Wen}, {and} \bibinfo{person}{Philip~S Yu}.} \bibinfo{year}{2023}\natexlab{}.
\newblock \showarticletitle{Automatic table union search with tabular representation learning}. In \bibinfo{booktitle}{\emph{Findings of the Association for Computational Linguistics: ACL 2023}}. \bibinfo{pages}{3786--3800}.
\newblock


\bibitem[Hui et~al\mbox{.}(2024)]%
        {hui2024uda}
\bibfield{author}{\bibinfo{person}{Yulong Hui}, \bibinfo{person}{Yao Lu}, {and} \bibinfo{person}{Huanchen Zhang}.} \bibinfo{year}{2024}\natexlab{}.
\newblock \showarticletitle{Uda: A benchmark suite for retrieval augmented generation in real-world document analysis}.
\newblock \bibinfo{journal}{\emph{Advances in Neural Information Processing Systems}}  \bibinfo{volume}{37} (\bibinfo{year}{2024}), \bibinfo{pages}{67200--67217}.
\newblock


\bibitem[Jiang et~al\mbox{.}(2023)]%
        {jiang2023mistral7b}
\bibfield{author}{\bibinfo{person}{Albert~Q. Jiang}, \bibinfo{person}{Alexandre Sablayrolles}, \bibinfo{person}{Arthur Mensch}, \bibinfo{person}{Chris Bamford}, \bibinfo{person}{Devendra~Singh Chaplot}, \bibinfo{person}{Diego de~las Casas}, \bibinfo{person}{Florian Bressand}, \bibinfo{person}{Gianna Lengyel}, \bibinfo{person}{Guillaume Lample}, \bibinfo{person}{Lucile Saulnier}, \bibinfo{person}{Lélio~Renard Lavaud}, \bibinfo{person}{Marie-Anne Lachaux}, \bibinfo{person}{Pierre Stock}, \bibinfo{person}{Teven~Le Scao}, \bibinfo{person}{Thibaut Lavril}, \bibinfo{person}{Thomas Wang}, \bibinfo{person}{Timothée Lacroix}, {and} \bibinfo{person}{William~El Sayed}.} \bibinfo{year}{2023}\natexlab{}.
\newblock \bibinfo{title}{Mistral 7B}.
\newblock
\newblock
\showeprint[arxiv]{2310.06825}~[cs.CL]
\urldef\tempurl%
\url{https://arxiv.org/abs/2310.06825}
\showURL{%
\tempurl}


\bibitem[Joshi et~al\mbox{.}(2017)]%
        {joshi2017triviaqa}
\bibfield{author}{\bibinfo{person}{Mandar Joshi}, \bibinfo{person}{Eunsol Choi}, \bibinfo{person}{Daniel~S Weld}, {and} \bibinfo{person}{Luke Zettlemoyer}.} \bibinfo{year}{2017}\natexlab{}.
\newblock \showarticletitle{Triviaqa: A large scale distantly supervised challenge dataset for reading comprehension}.
\newblock \bibinfo{journal}{\emph{arXiv preprint arXiv:1705.03551}} (\bibinfo{year}{2017}).
\newblock


\bibitem[Karpukhin et~al\mbox{.}(2020)]%
        {karpukhin2020dense}
\bibfield{author}{\bibinfo{person}{Vladimir Karpukhin}, \bibinfo{person}{Barlas Oguz}, \bibinfo{person}{Sewon Min}, \bibinfo{person}{Patrick Lewis}, \bibinfo{person}{Ledell Wu}, \bibinfo{person}{Sergey Edunov}, \bibinfo{person}{Danqi Chen}, {and} \bibinfo{person}{Wen-tau Yih}.} \bibinfo{year}{2020}\natexlab{}.
\newblock \showarticletitle{Dense Passage Retrieval for Open-Domain Question Answering}. In \bibinfo{booktitle}{\emph{Proceedings of the 2020 Conference on Empirical Methods in Natural Language Processing (EMNLP)}}. \bibinfo{publisher}{ACL}, \bibinfo{pages}{6769--6781}.
\newblock


\bibitem[Khattab and Zaharia(2020)]%
        {khattab2020colbert}
\bibfield{author}{\bibinfo{person}{Omar Khattab} {and} \bibinfo{person}{Matei Zaharia}.} \bibinfo{year}{2020}\natexlab{}.
\newblock \showarticletitle{Colbert: Efficient and effective passage search via contextualized late interaction over bert}. In \bibinfo{booktitle}{\emph{Proceedings of the 43rd International ACM SIGIR conference on research and development in Information Retrieval}}. \bibinfo{pages}{39--48}.
\newblock


\bibitem[Konan et~al\mbox{.}(2024)]%
        {konan2024automating}
\bibfield{author}{\bibinfo{person}{Sachin Konan}, \bibinfo{person}{Larry Rudolph}, {and} \bibinfo{person}{Scott Affens}.} \bibinfo{year}{2024}\natexlab{}.
\newblock \showarticletitle{Automating the Generation of a Functional Semantic Types Ontology with Foundational Models}. In \bibinfo{booktitle}{\emph{Proceedings of the 2024 Conference of the North American Chapter of the Association for Computational Linguistics: Human Language Technologies (Volume 6: Industry Track)}}. \bibinfo{pages}{248--265}.
\newblock


\bibitem[Kwiatkowski et~al\mbox{.}(2019)]%
        {kwiatkowski2019natural}
\bibfield{author}{\bibinfo{person}{Tom Kwiatkowski}, \bibinfo{person}{Jennimaria Palomaki}, \bibinfo{person}{Olivia Redfield}, \bibinfo{person}{Michael Collins}, \bibinfo{person}{Ankur Parikh}, \bibinfo{person}{Chris Alberti}, \bibinfo{person}{David Epstein}, \bibinfo{person}{Illia Polosukhin}, \bibinfo{person}{Jacob Devlin}, \bibinfo{person}{Kenton Lee}, \bibinfo{person}{Kristina Toutanova}, \bibinfo{person}{Llion Jones}, \bibinfo{person}{Matthew Kelcey}, \bibinfo{person}{Ming-Wei Chang}, \bibinfo{person}{Andrew Dai}, \bibinfo{person}{Jakob Uszkoreit}, \bibinfo{person}{Quoc Le}, {and} \bibinfo{person}{Slav Petrov}.} \bibinfo{year}{2019}\natexlab{}.
\newblock \showarticletitle{Natural Questions: A Benchmark for Question Answering Research}. In \bibinfo{booktitle}{\emph{Proceedings of the 2019 Conference on Empirical Methods in Natural Language Processing (EMNLP)}}. \bibinfo{publisher}{ACL}, \bibinfo{pages}{5358--5370}.
\newblock


\bibitem[Leek(2015)]%
        {datascience_book}
\bibfield{author}{\bibinfo{person}{Jeff Leek}.} \bibinfo{year}{2015}\natexlab{}.
\newblock \bibinfo{booktitle}{\emph{The Elements of Data Analytic Style}}.
\newblock \bibinfo{publisher}{Leanpub}.
\newblock
\urldef\tempurl%
\url{http://leanpub.com/datastyle}
\showURL{%
\tempurl}


\bibitem[Lei et~al\mbox{.}(2024)]%
        {lei2024spider}
\bibfield{author}{\bibinfo{person}{Fangyu Lei}, \bibinfo{person}{Jixuan Chen}, \bibinfo{person}{Yuxiao Ye}, \bibinfo{person}{Ruisheng Cao}, \bibinfo{person}{Dongchan Shin}, \bibinfo{person}{Hongjin Su}, \bibinfo{person}{Zhaoqing Suo}, \bibinfo{person}{Hongcheng Gao}, \bibinfo{person}{Wenjing Hu}, \bibinfo{person}{Pengcheng Yin}, {et~al\mbox{.}}} \bibinfo{year}{2024}\natexlab{}.
\newblock \showarticletitle{Spider 2.0: Evaluating language models on real-world enterprise text-to-sql workflows}.
\newblock \bibinfo{journal}{\emph{arXiv preprint arXiv:2411.07763}} (\bibinfo{year}{2024}).
\newblock


\bibitem[Li et~al\mbox{.}(2024)]%
        {li2024llms}
\bibfield{author}{\bibinfo{person}{Haitao Li}, \bibinfo{person}{Qian Dong}, \bibinfo{person}{Junjie Chen}, \bibinfo{person}{Huixue Su}, \bibinfo{person}{Yujia Zhou}, \bibinfo{person}{Qingyao Ai}, \bibinfo{person}{Ziyi Ye}, {and} \bibinfo{person}{Yiqun Liu}.} \bibinfo{year}{2024}\natexlab{}.
\newblock \showarticletitle{Llms-as-judges: a comprehensive survey on llm-based evaluation methods}.
\newblock \bibinfo{journal}{\emph{arXiv preprint arXiv:2412.05579}} (\bibinfo{year}{2024}).
\newblock


\bibitem[Li et~al\mbox{.}(2023)]%
        {li2023starcoder}
\bibfield{author}{\bibinfo{person}{Raymond Li}, \bibinfo{person}{Loubna~Ben Allal}, \bibinfo{person}{Yangtian Zi}, \bibinfo{person}{Niklas Muennighoff}, \bibinfo{person}{Denis Kocetkov}, \bibinfo{person}{Chenghao Mou}, \bibinfo{person}{Marc Marone}, \bibinfo{person}{Christopher Akiki}, \bibinfo{person}{Jia Li}, \bibinfo{person}{Jenny Chim}, {et~al\mbox{.}}} \bibinfo{year}{2023}\natexlab{}.
\newblock \showarticletitle{Starcoder: may the source be with you!}
\newblock \bibinfo{journal}{\emph{arXiv preprint arXiv:2305.06161}} (\bibinfo{year}{2023}).
\newblock


\bibitem[Liu et~al\mbox{.}(2023)]%
        {liu2023agentbench}
\bibfield{author}{\bibinfo{person}{Xiao Liu}, \bibinfo{person}{Hao Yu}, \bibinfo{person}{Hanchen Zhang}, \bibinfo{person}{Yifan Xu}, \bibinfo{person}{Xuanyu Lei}, \bibinfo{person}{Hanyu Lai}, \bibinfo{person}{Yu Gu}, \bibinfo{person}{Hangliang Ding}, \bibinfo{person}{Kaiwen Men}, \bibinfo{person}{Kejuan Yang}, {et~al\mbox{.}}} \bibinfo{year}{2023}\natexlab{}.
\newblock \showarticletitle{Agentbench: Evaluating llms as agents}.
\newblock \bibinfo{journal}{\emph{arXiv preprint arXiv:2308.03688}} (\bibinfo{year}{2023}).
\newblock


\bibitem[Luan et~al\mbox{.}(2021)]%
        {luan2021sparse}
\bibfield{author}{\bibinfo{person}{Yi Luan}, \bibinfo{person}{Jacob Eisenstein}, \bibinfo{person}{Kristina Toutanova}, {and} \bibinfo{person}{Michael Collins}.} \bibinfo{year}{2021}\natexlab{}.
\newblock \showarticletitle{Sparse, dense, and attentional representations for text retrieval}.
\newblock \bibinfo{journal}{\emph{Transactions of the Association for Computational Linguistics}}  \bibinfo{volume}{9} (\bibinfo{year}{2021}), \bibinfo{pages}{329--345}.
\newblock


\bibitem[Machado et~al\mbox{.}(2022)]%
        {machado2022data}
\bibfield{author}{\bibinfo{person}{In{\^e}s~Ara{\'u}jo Machado}, \bibinfo{person}{Carlos Costa}, {and} \bibinfo{person}{Maribel~Yasmina Santos}.} \bibinfo{year}{2022}\natexlab{}.
\newblock \showarticletitle{Data mesh: concepts and principles of a paradigm shift in data architectures}.
\newblock \bibinfo{journal}{\emph{Procedia Computer Science}}  \bibinfo{volume}{196} (\bibinfo{year}{2022}), \bibinfo{pages}{263--271}.
\newblock


\bibitem[Majumder et~al\mbox{.}(2024)]%
        {majumder2024discoverybench}
\bibfield{author}{\bibinfo{person}{Bodhisattwa~Prasad Majumder}, \bibinfo{person}{Harshit Surana}, \bibinfo{person}{Dhruv Agarwal}, \bibinfo{person}{Bhavana~Dalvi Mishra}, \bibinfo{person}{Abhijeetsingh Meena}, \bibinfo{person}{Aryan Prakhar}, \bibinfo{person}{Tirth Vora}, \bibinfo{person}{Tushar Khot}, \bibinfo{person}{Ashish Sabharwal}, {and} \bibinfo{person}{Peter Clark}.} \bibinfo{year}{2024}\natexlab{}.
\newblock \showarticletitle{Discoverybench: Towards data-driven discovery with large language models}.
\newblock \bibinfo{journal}{\emph{arXiv preprint arXiv:2407.01725}} (\bibinfo{year}{2024}).
\newblock


\bibitem[Mariscal et~al\mbox{.}(2010)]%
        {mariscal2010survey}
\bibfield{author}{\bibinfo{person}{Gonzalo Mariscal}, \bibinfo{person}{Oscar Marban}, {and} \bibinfo{person}{Covadonga Fernandez}.} \bibinfo{year}{2010}\natexlab{}.
\newblock \showarticletitle{A survey of data mining and knowledge discovery process models and methodologies}.
\newblock \bibinfo{journal}{\emph{The knowledge engineering review}} \bibinfo{volume}{25}, \bibinfo{number}{2} (\bibinfo{year}{2010}), \bibinfo{pages}{137--166}.
\newblock


\bibitem[Mialon et~al\mbox{.}(2023)]%
        {mialon2023gaia}
\bibfield{author}{\bibinfo{person}{Gr{\'e}goire Mialon}, \bibinfo{person}{Cl{\'e}mentine Fourrier}, \bibinfo{person}{Thomas Wolf}, \bibinfo{person}{Yann LeCun}, {and} \bibinfo{person}{Thomas Scialom}.} \bibinfo{year}{2023}\natexlab{}.
\newblock \showarticletitle{Gaia: a benchmark for general ai assistants}. In \bibinfo{booktitle}{\emph{The Twelfth International Conference on Learning Representations}}.
\newblock


\bibitem[Nan et~al\mbox{.}(2022)]%
        {nan2022fetaqa}
\bibfield{author}{\bibinfo{person}{Linyong Nan}, \bibinfo{person}{Chiachun Hsieh}, \bibinfo{person}{Ziming Mao}, \bibinfo{person}{Xi~Victoria Lin}, \bibinfo{person}{Neha Verma}, \bibinfo{person}{Rui Zhang}, \bibinfo{person}{Wojciech Kry{\'s}ci{\'n}ski}, \bibinfo{person}{Hailey Schoelkopf}, \bibinfo{person}{Riley Kong}, \bibinfo{person}{Xiangru Tang}, {et~al\mbox{.}}} \bibinfo{year}{2022}\natexlab{}.
\newblock \showarticletitle{FeTaQA: Free-form table question answering}.
\newblock \bibinfo{journal}{\emph{Transactions of the Association for Computational Linguistics}}  \bibinfo{volume}{10} (\bibinfo{year}{2022}), \bibinfo{pages}{35--49}.
\newblock


\bibitem[Nargesian et~al\mbox{.}(2018)]%
        {nargesian2018data}
\bibfield{author}{\bibinfo{person}{Fatemeh Nargesian}, \bibinfo{person}{Erkang Zhu}, \bibinfo{person}{Ren{\'e}e~J. Miller}, \bibinfo{person}{Ken~Q. Pu}, {and} \bibinfo{person}{Bahar Ghadiri~Bashardoost}.} \bibinfo{year}{2018}\natexlab{}.
\newblock \showarticletitle{Data Lake Organization}. In \bibinfo{booktitle}{\emph{Proceedings of the 2018 IEEE 34th International Conference on Data Engineering (ICDE)}}. \bibinfo{publisher}{IEEE}, \bibinfo{pages}{1033--1044}.
\newblock


\bibitem[Nguyen et~al\mbox{.}(2016)]%
        {nguyen2016msmarco}
\bibfield{author}{\bibinfo{person}{Tri Nguyen}, \bibinfo{person}{Mir Rosenberg}, \bibinfo{person}{Xia Song}, \bibinfo{person}{Jianfeng Gao}, \bibinfo{person}{Saurabh Tiwary}, \bibinfo{person}{Rangan Majumder}, {and} \bibinfo{person}{Li Deng}.} \bibinfo{year}{2016}\natexlab{}.
\newblock \showarticletitle{MS MARCO: A Human Generated MAchine Reading COmprehension Dataset}. In \bibinfo{booktitle}{\emph{Proceedings of the Workshop on Cognitive Computation: Integrating Neural and Symbolic Approaches (NeurIPS 2016)}}.
\newblock
\newblock
\shownote{Microsoft}.


\bibitem[Nogueira and Cho(2019)]%
        {nogueira2019passage}
\bibfield{author}{\bibinfo{person}{Rodrigo Nogueira} {and} \bibinfo{person}{Kyunghyun Cho}.} \bibinfo{year}{2019}\natexlab{}.
\newblock \showarticletitle{Passage Re-ranking with BERT}.
\newblock \bibinfo{journal}{\emph{arXiv preprint arXiv:1901.04085}} (\bibinfo{year}{2019}).
\newblock


\bibitem[Pasupat and Liang(2015)]%
        {WikiTableQuestions}
\bibfield{author}{\bibinfo{person}{Panupong Pasupat} {and} \bibinfo{person}{Percy Liang}.} \bibinfo{year}{2015}\natexlab{}.
\newblock \showarticletitle{Compositional Semantic Parsing on Semi-Structured Tables}. In \bibinfo{booktitle}{\emph{Proceedings of the 53rd Annual Meeting of the Association for Computational Linguistics and the 7th International Joint Conference on Natural Language Processing (Volume 1: Long Papers)}}. Association for Computational Linguistics, \bibinfo{address}{Beijing, China}, \bibinfo{pages}{1470--1480}.
\newblock
\urldef\tempurl%
\url{https://doi.org/10.3115/v1/P15-1142}
\showDOI{\tempurl}


\bibitem[Paton et~al\mbox{.}(2023)]%
        {paton2023dataset}
\bibfield{author}{\bibinfo{person}{Norman~W Paton}, \bibinfo{person}{Jiaoyan Chen}, {and} \bibinfo{person}{Zhenyu Wu}.} \bibinfo{year}{2023}\natexlab{}.
\newblock \showarticletitle{Dataset discovery and exploration: A survey}.
\newblock \bibinfo{journal}{\emph{Comput. Surveys}} \bibinfo{volume}{56}, \bibinfo{number}{4} (\bibinfo{year}{2023}), \bibinfo{pages}{1--37}.
\newblock


\bibitem[Pramanick et~al\mbox{.}(2024)]%
        {pramanick2024spiqa}
\bibfield{author}{\bibinfo{person}{Shraman Pramanick}, \bibinfo{person}{Rama Chellappa}, {and} \bibinfo{person}{Subhashini Venugopalan}.} \bibinfo{year}{2024}\natexlab{}.
\newblock \showarticletitle{Spiqa: A dataset for multimodal question answering on scientific papers}.
\newblock \bibinfo{journal}{\emph{Advances in Neural Information Processing Systems}}  \bibinfo{volume}{37} (\bibinfo{year}{2024}), \bibinfo{pages}{118807--118833}.
\newblock


\bibitem[Raja et~al\mbox{.}(2023)]%
        {raja2023icdar}
\bibfield{author}{\bibinfo{person}{Sachin Raja}, \bibinfo{person}{Ajoy Mondal}, {and} \bibinfo{person}{CV Jawahar}.} \bibinfo{year}{2023}\natexlab{}.
\newblock \showarticletitle{Icdar 2023 competition on visual question answering on business document images}. In \bibinfo{booktitle}{\emph{International Conference on Document Analysis and Recognition}}. Springer, \bibinfo{pages}{454--470}.
\newblock


\bibitem[Rajpurkar et~al\mbox{.}(2016)]%
        {rajpurkar2016squad}
\bibfield{author}{\bibinfo{person}{Pranav Rajpurkar}, \bibinfo{person}{Jian Zhang}, \bibinfo{person}{Konstantin Lopyrev}, {and} \bibinfo{person}{Percy Liang}.} \bibinfo{year}{2016}\natexlab{}.
\newblock \showarticletitle{Squad: 100,000+ questions for machine comprehension of text}.
\newblock \bibinfo{journal}{\emph{arXiv preprint arXiv:1606.05250}} (\bibinfo{year}{2016}).
\newblock


\bibitem[Raptis et~al\mbox{.}(2019)]%
        {raptis2019data}
\bibfield{author}{\bibinfo{person}{Theofanis~P Raptis}, \bibinfo{person}{Andrea Passarella}, {and} \bibinfo{person}{Marco Conti}.} \bibinfo{year}{2019}\natexlab{}.
\newblock \showarticletitle{Data management in industry 4.0: State of the art and open challenges}.
\newblock \bibinfo{journal}{\emph{Ieee Access}}  \bibinfo{volume}{7} (\bibinfo{year}{2019}), \bibinfo{pages}{97052--97093}.
\newblock


\bibitem[Ravat and Zhao(2019)]%
        {ravat2019data}
\bibfield{author}{\bibinfo{person}{Franck Ravat} {and} \bibinfo{person}{Yan Zhao}.} \bibinfo{year}{2019}\natexlab{}.
\newblock \showarticletitle{Data lakes: Trends and perspectives}. In \bibinfo{booktitle}{\emph{International Conference on Database and Expert Systems Applications}}. Springer, \bibinfo{pages}{304--313}.
\newblock


\bibitem[Robertson et~al\mbox{.}(2009)]%
        {robertson2009probabilistic}
\bibfield{author}{\bibinfo{person}{Stephen Robertson}, \bibinfo{person}{Hugo Zaragoza}, {et~al\mbox{.}}} \bibinfo{year}{2009}\natexlab{}.
\newblock \showarticletitle{The probabilistic relevance framework: BM25 and beyond}.
\newblock \bibinfo{journal}{\emph{Foundations and Trends{\textregistered} in Information Retrieval}} \bibinfo{volume}{3}, \bibinfo{number}{4} (\bibinfo{year}{2009}), \bibinfo{pages}{333--389}.
\newblock


\bibitem[Shraga et~al\mbox{.}(2020)]%
        {shraga2020web}
\bibfield{author}{\bibinfo{person}{Roee Shraga}, \bibinfo{person}{Haggai Roitman}, \bibinfo{person}{Guy Feigenblat}, {and} \bibinfo{person}{Mustafa Cannim}.} \bibinfo{year}{2020}\natexlab{}.
\newblock \showarticletitle{Web table retrieval using multimodal deep learning}. In \bibinfo{booktitle}{\emph{Proceedings of the 43rd international ACM SIGIR conference on research and development in information retrieval}}. \bibinfo{pages}{1399--1408}.
\newblock


\bibitem[Sicilia et~al\mbox{.}(2017)]%
        {sicilia2017community}
\bibfield{author}{\bibinfo{person}{Miguel-Angel Sicilia}, \bibinfo{person}{Elena Garc{\'\i}a-Barriocanal}, {and} \bibinfo{person}{Salvador S{\'a}nchez-Alonso}.} \bibinfo{year}{2017}\natexlab{}.
\newblock \showarticletitle{Community curation in open dataset repositories: insights from Zenodo}.
\newblock \bibinfo{journal}{\emph{Procedia Computer Science}}  \bibinfo{volume}{106} (\bibinfo{year}{2017}), \bibinfo{pages}{54--60}.
\newblock


\bibitem[Singhal et~al\mbox{.}(2001)]%
        {singhal2001modern}
\bibfield{author}{\bibinfo{person}{Amit Singhal} {et~al\mbox{.}}} \bibinfo{year}{2001}\natexlab{}.
\newblock \showarticletitle{Modern information retrieval: A brief overview}.
\newblock \bibinfo{journal}{\emph{IEEE Data Eng. Bull.}} \bibinfo{volume}{24}, \bibinfo{number}{4} (\bibinfo{year}{2001}), \bibinfo{pages}{35--43}.
\newblock


\bibitem[Song et~al\mbox{.}(2020)]%
        {song2020mpnet}
\bibfield{author}{\bibinfo{person}{Kaitao Song}, \bibinfo{person}{Xu Tan}, \bibinfo{person}{Tao Qin}, \bibinfo{person}{Jianfeng Lu}, {and} \bibinfo{person}{Tie-Yan Liu}.} \bibinfo{year}{2020}\natexlab{}.
\newblock \showarticletitle{Mpnet: Masked and permuted pre-training for language understanding}.
\newblock \bibinfo{journal}{\emph{Advances in neural information processing systems}}  \bibinfo{volume}{33} (\bibinfo{year}{2020}), \bibinfo{pages}{16857--16867}.
\newblock


\bibitem[Sparck~Jones(1972)]%
        {sparck1972statistical}
\bibfield{author}{\bibinfo{person}{Karen Sparck~Jones}.} \bibinfo{year}{1972}\natexlab{}.
\newblock \showarticletitle{A statistical interpretation of term specificity and its application in retrieval}.
\newblock \bibinfo{journal}{\emph{Journal of documentation}} \bibinfo{volume}{28}, \bibinfo{number}{1} (\bibinfo{year}{1972}), \bibinfo{pages}{11--21}.
\newblock


\bibitem[Wang et~al\mbox{.}({[n.\,d.]})]%
        {wangtoolbench}
\bibfield{author}{\bibinfo{person}{Guangyu Wang}, \bibinfo{person}{Jianhong Liu}, \bibinfo{person}{Meilin Zhou}, \bibinfo{person}{Xiaoming Chen}, \bibinfo{person}{Lihua Zhang}, {and} \bibinfo{person}{Zhihao Sun}.} \bibinfo{year}{[n.\,d.]}\natexlab{}.
\newblock \showarticletitle{ToolBench 2.0: Evaluating Long-Horizon and Multi-Step Tool Use in LLMs}.
\newblock  (\bibinfo{year}{[n.\,d.]}).
\newblock


\bibitem[Wang et~al\mbox{.}(2021)]%
        {wang2021milvus}
\bibfield{author}{\bibinfo{person}{Jianguo Wang}, \bibinfo{person}{Xiaomeng Yi}, \bibinfo{person}{Rentong Guo}, \bibinfo{person}{Hai Jin}, \bibinfo{person}{Peng Xu}, \bibinfo{person}{Shengjun Li}, \bibinfo{person}{Xiangyu Wang}, \bibinfo{person}{Xiangzhou Guo}, \bibinfo{person}{Chengming Li}, \bibinfo{person}{Xiaohai Xu}, {et~al\mbox{.}}} \bibinfo{year}{2021}\natexlab{}.
\newblock \showarticletitle{Milvus: A purpose-built vector data management system}. In \bibinfo{booktitle}{\emph{Proceedings of the 2021 international conference on management of data}}. \bibinfo{pages}{2614--2627}.
\newblock


\bibitem[Wang et~al\mbox{.}(2024c)]%
        {wang2024multilingual}
\bibfield{author}{\bibinfo{person}{Liang Wang}, \bibinfo{person}{Nan Yang}, \bibinfo{person}{Xiaolong Huang}, \bibinfo{person}{Linjun Yang}, \bibinfo{person}{Rangan Majumder}, {and} \bibinfo{person}{Furu Wei}.} \bibinfo{year}{2024}\natexlab{c}.
\newblock \showarticletitle{Multilingual E5 Text Embeddings: A Technical Report}.
\newblock \bibinfo{journal}{\emph{arXiv preprint arXiv:2402.05672}} (\bibinfo{year}{2024}).
\newblock


\bibitem[Wang et~al\mbox{.}(2024b)]%
        {wang2024large}
\bibfield{author}{\bibinfo{person}{Peiyi Wang}, \bibinfo{person}{Lei Li}, \bibinfo{person}{Liang Chen}, \bibinfo{person}{Zefan Cai}, \bibinfo{person}{Dawei Zhu}, \bibinfo{person}{Binghuai Lin}, \bibinfo{person}{Yunbo Cao}, \bibinfo{person}{Lingpeng Kong}, \bibinfo{person}{Qi Liu}, \bibinfo{person}{Tianyu Liu}, {et~al\mbox{.}}} \bibinfo{year}{2024}\natexlab{b}.
\newblock \showarticletitle{Large language models are not fair evaluators}. In \bibinfo{booktitle}{\emph{Proceedings of the 62nd Annual Meeting of the Association for Computational Linguistics (Volume 1: Long Papers)}}. \bibinfo{pages}{9440--9450}.
\newblock


\bibitem[Wang et~al\mbox{.}(2024a)]%
        {wang2024revisiting}
\bibfield{author}{\bibinfo{person}{Yuqi Wang}, \bibinfo{person}{Lyuhao Chen}, \bibinfo{person}{Songcheng Cai}, \bibinfo{person}{Zhijian Xu}, {and} \bibinfo{person}{Yilun Zhao}.} \bibinfo{year}{2024}\natexlab{a}.
\newblock \showarticletitle{Revisiting automated evaluation for long-form table question answering}. In \bibinfo{booktitle}{\emph{Proceedings of the 2024 Conference on Empirical Methods in Natural Language Processing}}. \bibinfo{pages}{14696--14706}.
\newblock


\bibitem[Wang et~al\mbox{.}(2023)]%
        {wang2023self}
\bibfield{author}{\bibinfo{person}{Yizhong Wang}, \bibinfo{person}{Yeganeh Kordi}, \bibinfo{person}{Swaroop Mishra}, \bibinfo{person}{Alisa Liu}, \bibinfo{person}{Noah~A Smith}, \bibinfo{person}{Daniel Khashabi}, {and} \bibinfo{person}{Hannaneh Hajishirzi}.} \bibinfo{year}{2023}\natexlab{}.
\newblock \showarticletitle{Self-instruct: Aligning language models with self-generated instructions}. In \bibinfo{booktitle}{\emph{Proceedings of the 61st annual meeting of the association for computational linguistics (volume 1: long papers)}}. \bibinfo{pages}{13484--13508}.
\newblock


\bibitem[Xu et~al\mbox{.}(2023)]%
        {xu2023wizardlm}
\bibfield{author}{\bibinfo{person}{Can Xu}, \bibinfo{person}{Qingfeng Sun}, \bibinfo{person}{Kai Zheng}, \bibinfo{person}{Xiubo Geng}, \bibinfo{person}{Pu Zhao}, \bibinfo{person}{Jiazhan Feng}, \bibinfo{person}{Chongyang Tao}, {and} \bibinfo{person}{Daxin Jiang}.} \bibinfo{year}{2023}\natexlab{}.
\newblock \showarticletitle{Wizardlm: Empowering large language models to follow complex instructions}.
\newblock \bibinfo{journal}{\emph{arXiv preprint arXiv:2304.12244}} (\bibinfo{year}{2023}).
\newblock


\bibitem[Yang et~al\mbox{.}(2025)]%
        {yang2025qwen3}
\bibfield{author}{\bibinfo{person}{An Yang}, \bibinfo{person}{Anfeng Li}, \bibinfo{person}{Baosong Yang}, \bibinfo{person}{Beichen Zhang}, \bibinfo{person}{Binyuan Hui}, \bibinfo{person}{Bo Zheng}, \bibinfo{person}{Bowen Yu}, \bibinfo{person}{Chang Gao}, \bibinfo{person}{Chengen Huang}, \bibinfo{person}{Chenxu Lv}, {et~al\mbox{.}}} \bibinfo{year}{2025}\natexlab{}.
\newblock \showarticletitle{Qwen3 technical report}.
\newblock \bibinfo{journal}{\emph{arXiv preprint arXiv:2505.09388}} (\bibinfo{year}{2025}).
\newblock


\bibitem[Yu et~al\mbox{.}(2023)]%
        {yu2023metamath}
\bibfield{author}{\bibinfo{person}{Longhui Yu}, \bibinfo{person}{Weisen Jiang}, \bibinfo{person}{Han Shi}, \bibinfo{person}{Jincheng Yu}, \bibinfo{person}{Zhengying Liu}, \bibinfo{person}{Yu Zhang}, \bibinfo{person}{James~T Kwok}, \bibinfo{person}{Zhenguo Li}, \bibinfo{person}{Adrian Weller}, {and} \bibinfo{person}{Weiyang Liu}.} \bibinfo{year}{2023}\natexlab{}.
\newblock \showarticletitle{Metamath: Bootstrap your own mathematical questions for large language models}.
\newblock \bibinfo{journal}{\emph{arXiv preprint arXiv:2309.12284}} (\bibinfo{year}{2023}).
\newblock


\bibitem[Zhao et~al\mbox{.}(2025)]%
        {zhao2025kalmembeddingv2}
\bibfield{author}{\bibinfo{person}{Xinping Zhao}, \bibinfo{person}{Xinshuo Hu}, \bibinfo{person}{Zifei Shan}, \bibinfo{person}{Shouzheng Huang}, \bibinfo{person}{Yao Zhou}, \bibinfo{person}{Xin Zhang}, \bibinfo{person}{Zetian Sun}, \bibinfo{person}{Zhenyu Liu}, \bibinfo{person}{Dongfang Li}, \bibinfo{person}{Xinyuan Wei}, \bibinfo{person}{Youcheng Pan}, \bibinfo{person}{Yang Xiang}, \bibinfo{person}{Meishan Zhang}, \bibinfo{person}{Haofen Wang}, \bibinfo{person}{Jun Yu}, \bibinfo{person}{Baotian Hu}, {and} \bibinfo{person}{Min Zhang}.} \bibinfo{year}{2025}\natexlab{}.
\newblock \bibinfo{title}{KaLM-Embedding-V2: Superior Training Techniques and Data Inspire A Versatile Embedding Model}.
\newblock
\newblock
\showeprint[arxiv]{2506.20923}~[cs.CL]
\urldef\tempurl%
\url{https://arxiv.org/abs/2506.20923}
\showURL{%
\tempurl}


\bibitem[Zheng et~al\mbox{.}(2023)]%
        {zheng2023judging}
\bibfield{author}{\bibinfo{person}{Lianmin Zheng}, \bibinfo{person}{Wei-Lin Chiang}, \bibinfo{person}{Ying Sheng}, \bibinfo{person}{Siyuan Zhuang}, \bibinfo{person}{Zhanghao Wu}, \bibinfo{person}{Yonghao Zhuang}, \bibinfo{person}{Zi Lin}, \bibinfo{person}{Zhuohan Li}, \bibinfo{person}{Dacheng Li}, \bibinfo{person}{Eric Xing}, {et~al\mbox{.}}} \bibinfo{year}{2023}\natexlab{}.
\newblock \showarticletitle{Judging llm-as-a-judge with mt-bench and chatbot arena}.
\newblock \bibinfo{journal}{\emph{Advances in neural information processing systems}}  \bibinfo{volume}{36} (\bibinfo{year}{2023}), \bibinfo{pages}{46595--46623}.
\newblock


\bibitem[Zhu et~al\mbox{.}(2022)]%
        {zhu2022towards}
\bibfield{author}{\bibinfo{person}{Fengbin Zhu}, \bibinfo{person}{Wenqiang Lei}, \bibinfo{person}{Fuli Feng}, \bibinfo{person}{Chao Wang}, \bibinfo{person}{Haozhou Zhang}, {and} \bibinfo{person}{Tat-Seng Chua}.} \bibinfo{year}{2022}\natexlab{}.
\newblock \showarticletitle{Towards complex document understanding by discrete reasoning}. In \bibinfo{booktitle}{\emph{Proceedings of the 30th ACM International Conference on Multimedia}}. \bibinfo{pages}{4857--4866}.
\newblock


\bibitem[Zhu et~al\mbox{.}(2021)]%
        {zhu2021tat}
\bibfield{author}{\bibinfo{person}{Fengbin Zhu}, \bibinfo{person}{Wenqiang Lei}, \bibinfo{person}{Youcheng Huang}, \bibinfo{person}{Chao Wang}, \bibinfo{person}{Shuo Zhang}, \bibinfo{person}{Jiancheng Lv}, \bibinfo{person}{Fuli Feng}, {and} \bibinfo{person}{Tat-Seng Chua}.} \bibinfo{year}{2021}\natexlab{}.
\newblock \showarticletitle{TAT-QA: A question answering benchmark on a hybrid of tabular and textual content in finance}.
\newblock \bibinfo{journal}{\emph{arXiv preprint arXiv:2105.07624}} (\bibinfo{year}{2021}).
\newblock


\end{thebibliography}


\clearpage

\onecolumn 
\appendix
\section{Prompts Used for Generation and Validation}
\label{app:prompts}
In this section, we provide details on the prompt design used for the LLM ensemble and a concrete case study illustrating the transformation from raw QA pairs to a Data Product Request (DPR).
\begin{promptbox}[Data Product Request Generation Prompt]
\textbf{Role:} You are a Data Product Request Generator.

\medskip
\textbf{Context:}  
A \emph{Data Product} is a self-contained, reusable, and consumable data asset designed to deliver specific value to its users for data-driven use cases.  
A \emph{Data Product Request (DPR)} is a high-level specification of what data and analysis the user needs.

\medskip
\textbf{Given:}  
A cluster containing multiple tables, where each table includes:
\begin{itemize}[nosep,leftmargin=*]
  \item a title (short description of the table);
  \item a list of column headers;
  \item a set of user questions that can be answered from the table.
\end{itemize}

\textbf{Task:}  
Write one DPR that effectively represents the combined data needs across all tables in the cluster.

\textbf{Instructions:}
\begin{itemize}[nosep,leftmargin=*]
  \item Do not copy or rephrase the input questions; integrate the tables’ information directly.
  \item Identify distinct analytical goals and write one sentence per goal.
  \item Use a clear, professional tone suitable for real-world data users.
  \item Focus on general insights, comparisons, relationships, or patterns that data engineers should support.
\end{itemize}

\textbf{Output format:} Return only the final DPR as plain text.

\textbf{Example DPRs:}
\begin{enumerate}[nosep,leftmargin=*]
  \item Gather data on hospital readmission rates for heart failure patients across different regions, and analyze which patient demographics or treatment protocols are most strongly associated with reduced readmission.
  \item Compile data on the highest-grossing films of the past 15 years and analyze how factors such as genre, director, production budget, and release season contribute to box office performance.
  \item Collect data showing changes in undergraduate admission rates at top U.S. public universities over the last decade, and assess how SAT scores, tuition, and diversity metrics influence these trends.
  \item Collect data that will allow queries on student and professor performance, including course satisfaction, grades, and demographics. It should also track student research capability, GPA, and course difficulty, as well as professor teaching ability and popularity.
  \item Compile a dataset that will allow queries on customer spending habits and behavior, including payment methods, consumption patterns, and average prices paid. It should also track spending changes over time and across locations (e.g., gas stations) to support segmentation and behavioral analysis.
\end{enumerate}
\end{promptbox}

\begin{promptbox}[Table Relevance Validator Prompt]
\textbf{Role:} You are a domain expert knowledgeable in building data products and data catalogues, with expertise in evaluating DPR quality.

\medskip
\textbf{Context:}  
A \emph{Data Product} is a self-contained, reusable, and consumable data asset designed to deliver specific value to its users for data-driven use cases.  
A \emph{Data Product Request (DPR)} is a high-level specification of what data and analysis the user needs.

\medskip
\textbf{Your task:}  
Given:
\begin{enumerate}[nosep,leftmargin=*]
  \item A data product request (DPR);
  \item A list of tables with each table’s title and columns.
\end{enumerate}

Identify any tables that are \textbf{not relevant} to the DPR.

\textbf{Definition of irrelevance:}
\begin{itemize}[nosep,leftmargin=*]
  \item A table is not relevant if it does not contain information useful to answer the DPR.
  \item A table need not perfectly match the DPR, but it should contain at least some overlapping or supporting information.
\end{itemize}

\textbf{Output:}
\begin{itemize}[nosep,leftmargin=*]
  \item List all tables identified as not relevant.
  \item For each, provide a short reasoning for the decision.
\end{itemize}
\end{promptbox}

\begin{promptbox}[DPR Quality Scoring Prompt (\texttt{LLMeval})]
\textbf{Role:} You are \texttt{LLMeval}, an expert evaluator of the quality of Data Product Requests (DPRs).

\medskip
\textbf{Context:}  
A \emph{Data Product} is a self-contained, reusable, and consumable data asset designed to deliver specific value to its users.  
A \emph{Data Product Request (DPR)} specifies what data and analysis the user needs.

\medskip
\textbf{Task:}  
Given a DPR, assign a score from \textbf{1} to \textbf{5} for each evaluation criterion below using the provided Likert scales.

\begin{enumerate}[leftmargin=*]
  \item \textbf{Quality — Level of Abstraction (1–5):}  
  Is the DPR phrased in a high-level, actionable manner suitable for guiding downstream data or analysis tasks?
  \begin{itemize}[nosep]
    \item 1 = Bad — Ambiguous or factoid; not actionable.
    \item 2 = Weak — Partially unclear or overly specific.
    \item 3 = Adequate — Generally appropriate but lacks full abstraction.
    \item 4 = Strong — Professional, abstract, and actionable with minor lapses.
    \item 5 = Ideal — Fully professional, concise, abstract, and actionable.
  \end{itemize}

  \item \textbf{Clarity (1–5):}  
  Is the DPR clearly and unambiguously written?
  \begin{itemize}[nosep]
    \item 1 = Unclear — Hard to understand or incomplete.
    \item 2 = Somewhat unclear — Several ambiguities.
    \item 3 = Moderately clear — Understandable with minor confusion.
    \item 4 = Clear — Easy to follow with rare issues.
    \item 5 = Crystal clear — Fluent and immediately understandable throughout.
  \end{itemize}
\end{enumerate}
\end{promptbox}

\section{Pipeline workflow}

\subsection{Case Study: From Table QA to Data Product Request.}
Figure~\ref{fig:framework} illustrates the high-level pipeline. Here, we describe a specific instance from the Finance domain.

We illustrate our framework \framework through a detailed walkthrough using source material from the HybridQA dataset~\cite{chen2020hybridqa}. This example demonstrates how our framework transforms granular question-answer pairs into comprehensive data product requests that integrate multiple structured and unstructured information sources.

\noindent\textbf{Source Material.}
The starting point is a Wikipedia table from the entry \emph{``List of UEFA European Championship winning managers.''} This table systematically records managerial achievements in European football championships with the schema: \texttt{[Year, Winning Manager, Country, Score in Final, Venue, Opponent]}. For instance, one row documents that in 2012, Vicente del Bosque (Spain) defeated Italy 4--0 in the final held in Kyiv. 

The original HybridQA dataset associates this table with multiple fine-grained questions, each targeting a specific cell or relationship:
\begin{itemize}
    \item \emph{``Who was the manager of the winning team in 2008?''}
    \item \emph{``Which country did Vicente del Bosque manage when he won the title?''}
    \item \emph{``What was the score in the 2012 final?''}
    \item \emph{``Where was the 1984 championship final held?''}
\end{itemize}
Each question is paired with a supporting passage extracted from the linked Wikipedia article. For example, the passage for the 2012 question states: \emph{``Spain, managed by Vicente del Bosque, defeated Italy 4--0 in the 2012 UEFA European Championship final, held at the Olympic Stadium in Kyiv. This victory marked Spain's third European Championship title.''} These passages provide narrative context beyond what appears in the structured table cells.

\noindent\textbf{Topic Clustering via Semantic Embeddings.}
To identify thematically related tables across the dataset, we first construct a semantic representation of each table. Specifically, we concatenate all questions associated with a table and embed the resulting text using a pre-trained sentence encoder. This embedding captures the collective semantic intent expressed across all QA pairs for that table.

For the UEFA managers table, the concatenated questions emphasize themes around ``tournament winners,'' ``national team managers,'' ``championship finals,'' and ``historical results.'' When we perform hierarchical clustering over all table embeddings in HybridQA, this table naturally groups with other tournament-focused tables, including:
\begin{itemize}
    \item \emph{``List of FIFA World Cup winning managers''} (schema: \texttt{[Year, Manager, Country, Host Nation, Score]})
    \item \emph{``List of Copa América winning managers''} (schema: \texttt{[Year, Manager, Country, Venue, Final Score]})
    \item \emph{``African Cup of Nations winning managers''} (schema: \texttt{[Year, Manager, Country, Final Result]})
\end{itemize}
The clustering algorithm reveals that these tables, despite originating from different Wikipedia articles, share a common analytical focus: documenting managerial success across major international football competitions. This semantic grouping reflects a realistic data analyst's workflow, where relevant information is scattered across multiple sources that must be discovered and integrated.

Notably, the clustering also excludes tables that superficially mention football but serve different purposes—such as tables listing \emph{``UEFA Champions League attendance records''} or \emph{``Player transfer fees by season.''} These are filtered because their questions emphasize different analytical dimensions (attendance patterns, financial transactions) rather than managerial achievements.

\noindent\textbf{Schema Refinement and Structural Alignment.}
Within the identified cluster, we perform a second stage of refinement based on schema compatibility. Not all semantically related tables are structurally compatible for joint analysis. We compute schema similarity by comparing column names and types, retaining only tables that share core attributes necessary for the analytical task.

In this example, the three tournament manager tables share a compatible structural pattern:
\begin{itemize}
    \item Core temporal dimension: \texttt{Year}
    \item Core entity: \texttt{Winning Manager}, \texttt{Manager}
    \item Geographic dimension: \texttt{Country}
    \item Outcome measure: \texttt{Score in Final}, \texttt{Final Score}
    \item Context: \texttt{Venue}, \texttt{Host Nation}
\end{itemize}
Although column names vary slightly (e.g., ``Score in Final'' vs. ``Final Score''), the semantic roles align, enabling cross-table analysis. A table describing \emph{``Player statistics by tournament''} would be excluded at this stage despite topical relevance, as its schema (\texttt{[Player Name, Goals, Assists, Minutes Played]}) does not align with the managerial focus.

For each retained table, we also aggregate all linked passages. In this case, the data product includes passages describing:
\begin{itemize}
    \item Historical context for each tournament (e.g., ``The 2008 UEFA Championship was held in Austria and Switzerland...'')
    \item Notable managerial achievements (e.g., ``Vicente del Bosque became only the third manager to win both the World Cup and European Championship'')
    \item Details of specific finals that cannot be fully captured in tabular format
\end{itemize}
This aggregation ensures the data product contains both structured data (tables) and unstructured context (passages), mirroring real-world analytical workflows where both types of evidence are consulted.

\noindent\textbf{Data Product Request Generation via LLM Ensemble.}
Given the refined data product—three aligned tables and their associated passages—we now generate a professional-level analytical request that captures the collective information need. We employ an ensemble of three LLMs (GPT-4, Claude, and Llama-3) to avoid model-specific biases.

The prompt provides:
\begin{itemize}
    \item A summary of all table schemas with example rows
    \item Representative questions from each table
    \item Key themes extracted from passage content
    \item Instructions to formulate a realistic analyst request that requires all tables
\end{itemize}

For this data product, the three models independently generate candidate DPRs. After filtering for quality (coherence, specificity, analytical intent), the validated DPR selected by majority agreement is:
\begin{quote}
\emph{``Collect comprehensive data on major international football tournaments, including the UEFA European Championship, FIFA World Cup, and Copa América, with information on each year's winning manager, country, match outcome, and venue. The goal is to analyze patterns in managerial success across different competitions and time periods, identifying managers who achieved success in multiple tournaments and examining how home-field advantage and regional factors influence championship outcomes.''}
\end{quote}

This DPR exhibits several key properties:
\begin{enumerate}
    \item \textbf{Scope generalization}: It synthesizes 47 fine-grained QA pairs (across three tables) into a single analytical goal
    \item \textbf{Multi-table requirement}: Answering requires integrating all three tournament tables—no single table suffices
    \item \textbf{Analytical depth}: It goes beyond simple fact lookup to specify comparative analysis (``patterns,'' ``multiple tournaments,'' ``regional factors'')
    \item \textbf{Mixed evidence}: It implicitly requires both structured data (tables) and unstructured context (passages about venues, historical significance)
\end{enumerate}

Alternative candidate DPRs generated by the ensemble included variations emphasizing different analytical angles (e.g., temporal trends in scoring patterns, geographic distribution of winning nations). The final DPR was selected based on its balance of specificity and generality, ensuring it is neither too narrow (addressable by a single table) nor too vague (underspecified analytical intent).

\noindent\textbf{Final Benchmark Artifact.}  
The resulting benchmark entry consists of:  
(i) the validated DPR above,  
(ii) the corresponding set of tables—here, those describing winning managers across multiple tournaments,  
and (iii) the linked passages providing textual explanations of the outcomes.  
Each component maintains full provenance to the original HybridQA sources. 

To validate the benchmark quality, we apply three checks:
\begin{enumerate}
    \item \textbf{Necessity}: Human annotators verify that all three tables are required—removing any single table leaves key analytical questions unanswerable
    \item \textbf{Sufficiency}: The table and passage set contains all information needed to address the DPR
    \item \textbf{Realism}: Domain experts (data analysts with sports analytics experience) confirm the DPR reflects a plausible real-world analytical task
\end{enumerate}

This example shows how our pipeline transforms fragmented table-level QA pairs into integrated data product specifications. The resulting benchmark entry reflects realistic complexity: it requires discovering multiple relevant tables from a large corpus, understanding that they address a shared analytical goal despite structural variations, and recognizing that both structured and unstructured evidence contribute to a complete data product. This stands in contrast to traditional table QA benchmarks, where each question targets a single table and success is measured by cell-level accuracy rather than the ability to assemble coherent multi-table analytical resources.

\section{Human Evaluation Analysis}
\label{app:human_eval}

To ensure the reliability of our validation protocol, we conducted a human evaluation of the generated DPRs. We recruited six expert annotators to grade the requests across four dimensions: \textit{Quality}, \textit{Clarity}, \textit{Relevance}, and \textit{Completeness}.

Table~\ref{tab:annotation_by_annotator} presents the mean scores for each annotator ($n=51$ samples per annotator). The results indicate high inter-annotator consistency for \textit{Quality} and \textit{Relevance}. We observe slightly higher variance in \textit{Completeness}, likely due to the subjective nature of determining whether a request "fully" encompasses all potential insights in a large table. Despite this, the high scores for Quality ($>0.9$ for most annotators) validate the efficacy of our LLM ensemble in generating professional-grade requests.

\begin{table}[h!]
\centering
\caption{Mean score per annotator ($n = 51$ per annotator) across four quality dimensions.}
\label{tab:annotation_by_annotator}
\resizebox{0.48\textwidth}{!}{%
\begin{tabular}{lcccc}
\toprule
\textbf{Annotator} & \textbf{Quality} & \textbf{Clarity} & \textbf{Relevance} & \textbf{Completeness} \\
\midrule
A & 0.961 & 0.961 & 0.972 & 0.804 \\
B & 0.941 & 0.902 & 0.990 & 0.863 \\
C & 0.765 & 0.765 & 0.956 & 0.549 \\
D & 0.784 & 0.863 & 0.915 & 0.431 \\
E & 0.902 & 0.961 & 0.863 & 0.569 \\
F & 0.922 & 0.922 & 0.984 & 0.627 \\
\bottomrule
\end{tabular}}
\end{table}

\section{Extended Experimental Results}
\label{app:additional_results}

We performed extensive ablation studies to evaluate the impact of different model architectures and embedding strategies on retrieval performance.

\subsection{Impact of LLM Choice on Query Expansion}
Table~\ref{tab:results-table} investigates whether larger Language Models (LLMs) significantly improve the retrieval of Data Products when used for query expansion (DPR generation). We compare a baseline (no expansion) against expansions generated by \texttt{GPT-OSS-120B} and \texttt{Qwen3-VL-235B}.

Interestingly, we observe that retrieval performance saturates; the 235B parameter model yields performance comparable to the 120B model. In some settings, expansions can even degrade performance. This suggests that the \textit{quality} of the prompt engineering and the initial clustering (Stage 1 \& 2 of DPForge) is more critical than the sheer scale of the LLM used for generation.

\begin{table*}[!h]
\centering
\setlength{\tabcolsep}{4pt}
\renewcommand{\arraystretch}{1.02}
\caption{Retrieval performance using different LLMs to expand the DPR prior to retrieval for DP-HybridQA dataset. Results show Full Recall@100 remains stable across model sizes. All experiments use granite-125m-english embeddings.}
\label{tab:results-table}
\resizebox{\textwidth}{!}{%
\begin{tabular}{l *{16}{c}}
\toprule
\multirow{3}{*}{\textbf{LLM}} &
\multicolumn{6}{c}{\textbf{Tables}} &
\multicolumn{6}{c}{\textbf{Text}} &
\multicolumn{3}{c}{\textbf{Data Product}} \\
\cmidrule(lr){2-7}\cmidrule(lr){8-13}\cmidrule(lr){14-16}
& \multicolumn{3}{c}{\textbf{Recall@20}} &
\multicolumn{3}{c}{\textbf{Recall@100}} &
\multicolumn{3}{c}{\textbf{Recall@20}} &
\multicolumn{3}{c}{\textbf{Recall@100}} &
\multicolumn{3}{c}{\textbf{Full Recall@100}} \\
\cmidrule(lr){2-4}\cmidrule(lr){5-7}\cmidrule(lr){8-10}\cmidrule(lr){11-13}\cmidrule(lr){14-16}
& BM25 & Dense & Hybrid & BM25 & Dense & Hybrid & BM25 & Dense & Hybrid & BM25 & Dense & Hybrid & BM25 & Dense & Hybrid \\
\midrule
No expansion & 0.03 & 0.44 & 0.49 & 0.62 & 0.69 & 0.71 & 0.06 & 0.25 & 0.28 & 0.36 & 0.38 & 0.49 & 0.07 & 0.11 & 0.16 \\
GPT-OSS-120B & 0.03 & 0.43 & 0.47 & 0.61 & 0.68 & 0.71 & 0.06 & 0.25 & 0.27 & 0.36 & 0.38 & 0.49 & 0.07 & 0.11 & 0.16 \\
Qwen3-VL-235B & 0.03 & 0.42 & 0.45 & 0.58 & 0.67 & 0.69 & 0.06 & 0.25 & 0.27 & 0.34 & 0.37 & 0.47 & 0.06 & 0.10 & 0.15 \\
\bottomrule
\end{tabular}}
\end{table*}

\subsection{Embedding Model Benchmarks}
We further benchmark various embedding models to validate the retrieval component of \benchmark. Tables~\ref{tab:results-embeddings-tatqa} and~\ref{tab:results-embeddings-convfinqa} detail the performance on the DP-TATQA and DP-ConvFinQA subsets, respectively.

\textbf{DP-TATQA Results:} As shown in Table~\ref{tab:results-embeddings-tatqa}, larger embedding models such as \texttt{Qwen3} (8B) and \texttt{KaLM-Gemma3} (12B) significantly outperform standard baselines like \texttt{MPNet} (110M), particularly in Dense retrieval settings where semantic understanding of complex table headers is required.

\textbf{DP-ConvFinQA Results:} Table~\ref{tab:results-embeddings-convfinqa} demonstrates higher overall baseline scores due to the conversational nature of the dataset. Here, the gap between small and large models narrows, though \texttt{KaLM-Gemma3} still achieves state-of-the-art results with a Hybrid Full Recall@100 of 0.96.

\begin{table*}[!h]
\centering
\setlength{\tabcolsep}{4pt}
\renewcommand{\arraystretch}{1.02}
\caption{Retrieval performance across different embedding models on the \textbf{DP-TATQA} dataset.}
\label{tab:results-embeddings-tatqa}
\resizebox{\textwidth}{!}{%
\begin{tabular}{l c c *{11}{c}}
\toprule
\multirow{3}{*}{\textbf{Embedding Model}} &
\multirow{3}{*}{\parbox{1cm}{\centering\textbf{Model \\ Params.}}} &
\multirow{3}{*}{\parbox{1cm}{\centering\textbf{Embed \\ Dims.}}} &
\multicolumn{4}{c}{\textbf{Tables}} &
\multicolumn{4}{c}{\textbf{Text}} &
\multicolumn{2}{c}{\textbf{Data Product}} \\
\cmidrule(lr){4-7}\cmidrule(lr){8-11}\cmidrule(lr){12-13}
& & &
\multicolumn{2}{c}{\textbf{Recall@20}} &
\multicolumn{2}{c}{\textbf{Recall@100}} &
\multicolumn{2}{c}{\textbf{Recall@20}} &
\multicolumn{2}{c}{\textbf{Recall@100}} &
\multicolumn{2}{c}{\textbf{Full Recall@100}} \\
\cmidrule(lr){4-5}\cmidrule(lr){6-7}\cmidrule(lr){8-9}\cmidrule(lr){10-11}\cmidrule(lr){12-13}
& & &
Dense & Hybrid & Dense & Hybrid &
Dense & Hybrid & Dense & Hybrid &
Dense & Hybrid \\
\midrule
all-mpnet-base-v2~\cite{song2020mpnet} & 110M & 768 & 0.32 & 0.54 & 0.62 & 0.79 & 0.25 & 0.37 & 0.54 & 0.71 & 0.10 & 0.35 \\
granite-125m-english~\cite{awasthy2025graniteembeddingmodels} & 125M & 768 & 0.33 & 0.47 & 0.59 & 0.79 & 0.28 & 0.42 & 0.61 & 0.72 & 0.18 & 0.34 \\
multilingual-e5-large-instruct~\cite{wang2024multilingual} & 560M & 1024 & 0.24 & 0.43 & 0.55 & 0.78 & 0.29 & 0.40 & 0.55 & 0.69 & 0.09 & 0.34 \\
Qwen3~\cite{yang2025qwen3} & 8B & 4096 & 0.53 & 0.59 & 0.84 & 0.88 & 0.42 & 0.47 & 0.75 & 0.79 & 0.38 & 0.45 \\
KaLM-Gemma3-12B-2511~\cite{zhao2025kalmembeddingv2} & 12B & 3840 & 0.44 & 0.56 & 0.77 & 0.83 & 0.41 & 0.46 & 0.75 & 0.78 & 0.34 & 0.47 \\
\bottomrule
\end{tabular}}
\end{table*}

\begin{table*}[!h]
\centering
\setlength{\tabcolsep}{4pt}
\renewcommand{\arraystretch}{1.02}
\caption{Retrieval performance across different embedding models on the \textbf{DP-ConvFinQA} dataset.}
\label{tab:results-embeddings-convfinqa}
\resizebox{\textwidth}{!}{%
\begin{tabular}{l c c *{11}{c}}
\toprule
\multirow{3}{*}{\textbf{Embedding Model}} &
\multirow{3}{*}{\parbox{1cm}{\centering\textbf{Model \\ Params.}}} &
\multirow{3}{*}{\parbox{1cm}{\centering\textbf{Embed \\ Dims.}}} &
\multicolumn{4}{c}{\textbf{Tables}} &
\multicolumn{4}{c}{\textbf{Text}} &
\multicolumn{2}{c}{\textbf{Data Product}} \\
\cmidrule(lr){4-7}\cmidrule(lr){8-11}\cmidrule(lr){12-13}
& & &
\multicolumn{2}{c}{\textbf{Recall@20}} &
\multicolumn{2}{c}{\textbf{Recall@100}} &
\multicolumn{2}{c}{\textbf{Recall@20}} &
\multicolumn{2}{c}{\textbf{Recall@100}} &
\multicolumn{2}{c}{\textbf{Full Recall@100}} \\
\cmidrule(lr){4-5}\cmidrule(lr){6-7}\cmidrule(lr){8-9}\cmidrule(lr){10-11}\cmidrule(lr){12-13}
& & &
Dense & Hybrid & Dense & Hybrid &
Dense & Hybrid & Dense & Hybrid &
Dense & Hybrid \\
\midrule
all-mpnet-base-v2~\cite{song2020mpnet} & 110M & 768 & 0.81 & 0.92 & 0.94 & 0.97 & 0.82 & 0.91 & 0.92 & 0.98 & 0.85 & 0.94 \\
granite-125m-english~\cite{awasthy2025graniteembeddingmodels} & 125M & 768 & - & - & - & - & - & - & - & - & - & - \\
multilingual-e5-large-instruct~\cite{wang2024multilingual} & 560M & 1024 & 0.88 & 0.92 & 0.95 & 0.97 & 0.86 & 0.92 & 0.95 & 0.98 & 0.89 & 0.93 \\
Qwen3~\cite{yang2025qwen3} & 8B & 4096 & 0.93 & 0.94 & 0.98 & 0.98 & 0.91 & 0.94 & 0.97 & 0.98 & 0.93 & 0.93 \\
KaLM-Gemma3-12B-2511~\cite{zhao2025kalmembeddingv2} & 12B & 3840 & 0.93 & 0.95 & 0.99 & 0.99 & 0.92 & 0.94 & 0.98 & 0.99 & 0.95 & 0.96 \\
\bottomrule
\end{tabular}}
\end{table*}

\section{Benchmark Characteristics}
\label{app:dataset_stats}

\begin{figure}[h]
    \centering
    \begin{subfigure}{0.45\linewidth}
        \centering
        \includegraphics[width=\linewidth]{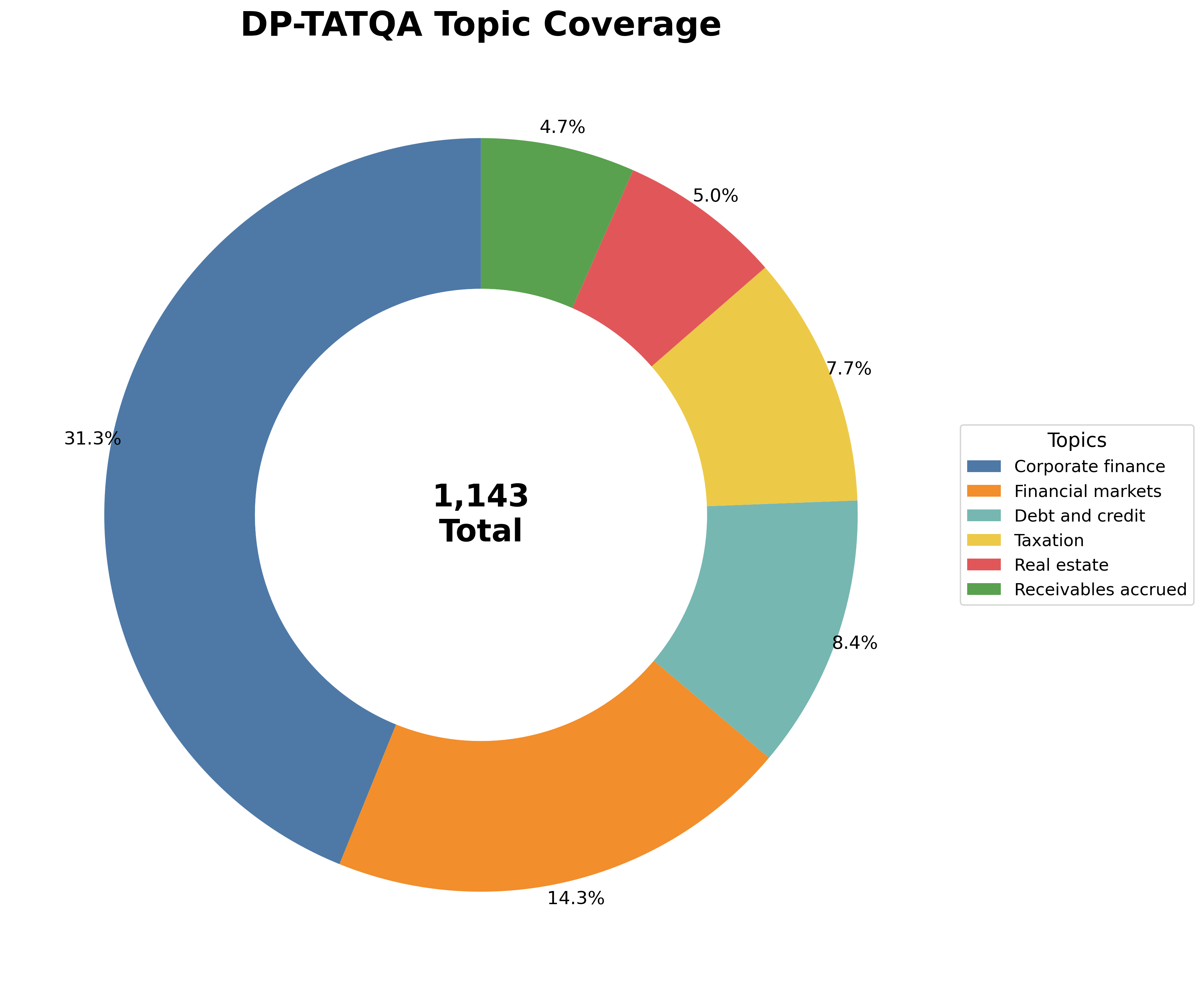}
        \caption{DP-TATQA}
        \label{fig:tatqa}
    \end{subfigure}\hfill
    \begin{subfigure}{0.45\linewidth}
        \centering
        \includegraphics[width=\linewidth]{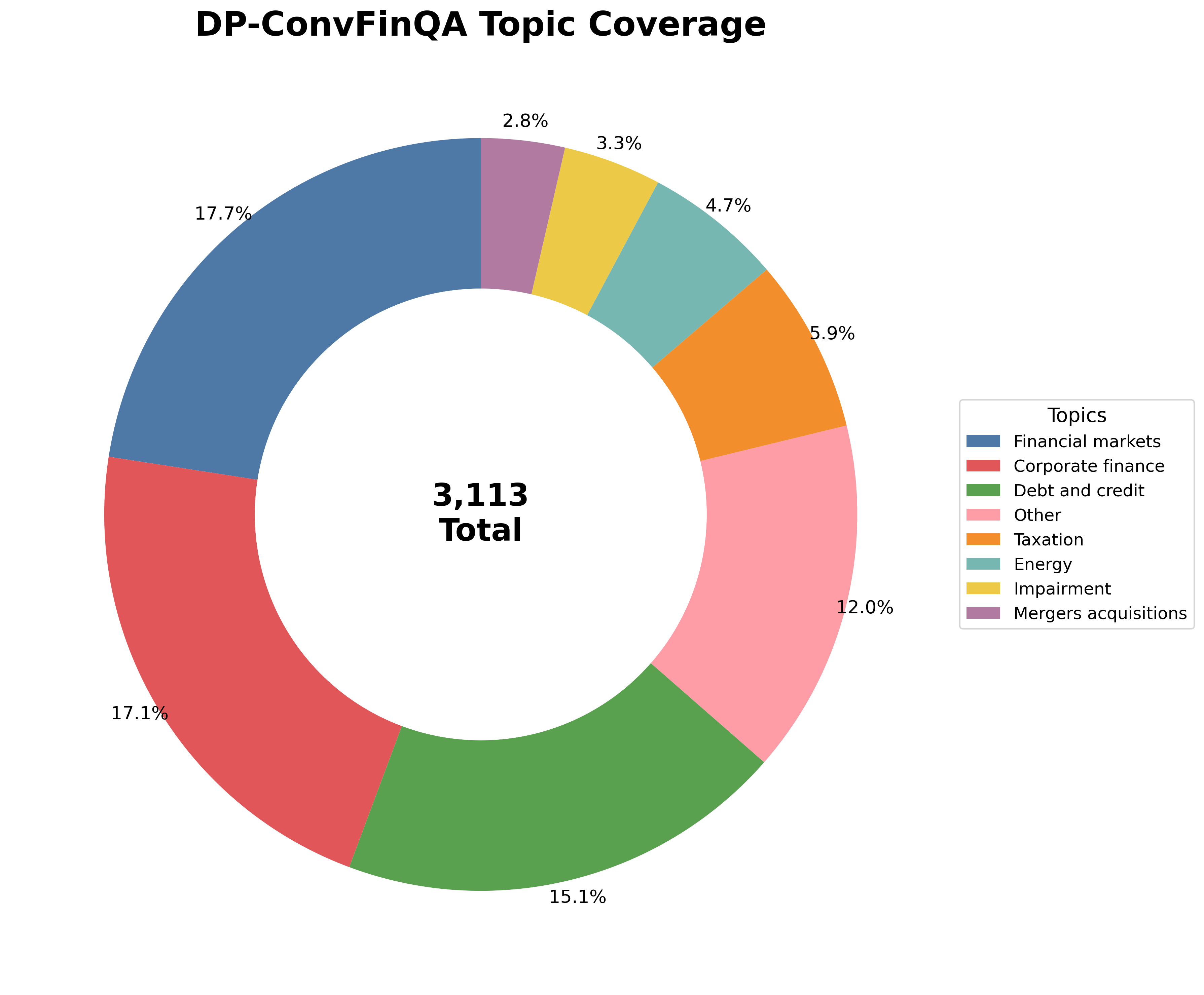}
        \caption{DP-ConvFinQA}
        \label{fig:convfinqa}
    \end{subfigure}
    \caption{Topic distributions of financial datasets. The clustering reveals distinct analytical domains suitable for data product discovery.}
    \label{fig:financial-topics}
\end{figure}

Finally, we visualize the topic distribution of the generated benchmark in Figure~\ref{fig:financial-topics}. The charts confirm that \benchmark maintains a diverse coverage of financial domains, inheriting the complex distribution of the source datasets (DP-TATQA and DP-ConvFinQA) while restructuring them into product-centric clusters.

\end{document}